\documentclass[submitting]{nst}
\usepackage{booktabs}
\usepackage{diagbox}
\usepackage{hyperref}
\usepackage[table,xcdraw]{xcolor}
\usepackage[normalem]{ulem}
\usepackage{makecell}
\usepackage{graphicx}
\usepackage{float}
\usepackage{amsmath}
\usepackage{subfigure,dcolumn}
\usepackage[T2A,T1]{fontenc}
\usepackage[russian,english]{babel} 

\usepackage{listings}
\begin{document}

\title{Design of a LYSO Crystal Electromagnetic Calorimeter for DarkSHINE Experiment}\thanks{This work was supported by National Key R\&D Program of China (Grant No.: 2023YFA1606904 and 2023YFA1606900), National Natural Science Foundation of China (Grant No.: 12150006), Shanghai Pilot Program for Basic Research—Shanghai Jiao Tong University (Grant No.: 21TQ1400209), and National Center for High-Level Talent Training in Mathematics, Physics, Chemistry, and Biology.}

\author{Zhiyu Zhao}\thanks{These authors contributed equally to this work.}
\affiliation{Tsung-Dao Lee Institute, Shanghai Jiao Tong University, 1 Lisuo Road, Shanghai 201210, China}
\affiliation{Institute of Nuclear and Particle Physics, School of Physics and Astronomy, 800 Dongchuan Road, Shanghai 200240, China}
\affiliation{Key Laboratory for Particle Astrophysics and Cosmology (MOE), Shanghai Key Laboratory for Particle Physics and Cosmology (SKLPPC), Shanghai Jiao Tong University, 800 Dongchuan Road, Shanghai 200240, China}

\author{Qibin Liu}\thanks{These authors contributed equally to this work.}
\affiliation{Tsung-Dao Lee Institute, Shanghai Jiao Tong University, 1 Lisuo Road, Shanghai 201210, China}
\affiliation{Institute of Nuclear and Particle Physics, School of Physics and Astronomy, 800 Dongchuan Road, Shanghai 200240, China}
\affiliation{Key Laboratory for Particle Astrophysics and Cosmology (MOE), Shanghai Key Laboratory for Particle Physics and Cosmology (SKLPPC), Shanghai Jiao Tong University, 800 Dongchuan Road, Shanghai 200240, China}

\author{Jiyuan Chen}
\affiliation{Institute of Nuclear and Particle Physics, School of Physics and Astronomy, 800 Dongchuan Road, Shanghai 200240, China}
\affiliation{Key Laboratory for Particle Astrophysics and Cosmology (MOE), Shanghai Key Laboratory for Particle Physics and Cosmology (SKLPPC), Shanghai Jiao Tong University, 800 Dongchuan Road, Shanghai 200240, China}
\affiliation{Tsung-Dao Lee Institute, Shanghai Jiao Tong University, 1 Lisuo Road, Shanghai 201210, China}

\author{Jing Chen}
\affiliation{Institute of Nuclear and Particle Physics, School of Physics and Astronomy, 800 Dongchuan Road, Shanghai 200240, China}
\affiliation{Key Laboratory for Particle Astrophysics and Cosmology (MOE), Shanghai Key Laboratory for Particle Physics and Cosmology (SKLPPC), Shanghai Jiao Tong University, 800 Dongchuan Road, Shanghai 200240, China}
\affiliation{Tsung-Dao Lee Institute, Shanghai Jiao Tong University, 1 Lisuo Road, Shanghai 201210, China}

\author{Junfeng Chen}
\affiliation{Center of Materials Science and Optoelectronics Engineering, University of Chinese Academy of Science, Beijing, 100049 China}
\affiliation{Shanghai Institute of Ceramics, Chinese Academy of Sciences, Shanghai, 201899 China}

\author{Xiang Chen}
\affiliation{Institute of Nuclear and Particle Physics, School of Physics and Astronomy, 800 Dongchuan Road, Shanghai 200240, China}
\affiliation{Key Laboratory for Particle Astrophysics and Cosmology (MOE), Shanghai Key Laboratory for Particle Physics and Cosmology (SKLPPC), Shanghai Jiao Tong University, 800 Dongchuan Road, Shanghai 200240, China}

\author{Changbo Fu}
\affiliation{Key Laboratory of Nuclear Physics and Ion-beam Application (MOE), Fudan University, Shanghai, 200443, China}
\affiliation{Institute of Modern Physics, Fudan University, Shanghai, 200443, China}

\author{Jun Guo}
\affiliation{Institute of Nuclear and Particle Physics, School of Physics and Astronomy, 800 Dongchuan Road, Shanghai 200240, China}
\affiliation{Key Laboratory for Particle Astrophysics and Cosmology (MOE), Shanghai Key Laboratory for Particle Physics and Cosmology (SKLPPC), Shanghai Jiao Tong University, 800 Dongchuan Road, Shanghai 200240, China}

\author{Kim Siang Khaw}
\affiliation{Tsung-Dao Lee Institute, Shanghai Jiao Tong University, 1 Lisuo Road, Shanghai 201210, China}
\affiliation{Institute of Nuclear and Particle Physics, School of Physics and Astronomy, 800 Dongchuan Road, Shanghai 200240, China}
\affiliation{Key Laboratory for Particle Astrophysics and Cosmology (MOE), Shanghai Key Laboratory for Particle Physics and Cosmology (SKLPPC), Shanghai Jiao Tong University, 800 Dongchuan Road, Shanghai 200240, China}

\author{Liang Li}
\affiliation{Institute of Nuclear and Particle Physics, School of Physics and Astronomy, 800 Dongchuan Road, Shanghai 200240, China}
\affiliation{Key Laboratory for Particle Astrophysics and Cosmology (MOE), Shanghai Key Laboratory for Particle Physics and Cosmology (SKLPPC), Shanghai Jiao Tong University, 800 Dongchuan Road, Shanghai 200240, China}

\author{Shu Li}
\email[Corresponding author, ]{Shu Li, shuli@sjtu.edu.cn}{}
\affiliation{Tsung-Dao Lee Institute, Shanghai Jiao Tong University, 1 Lisuo Road, Shanghai 201210, China}
\affiliation{Institute of Nuclear and Particle Physics, School of Physics and Astronomy, 800 Dongchuan Road, Shanghai 200240, China}
\affiliation{Key Laboratory for Particle Astrophysics and Cosmology (MOE), Shanghai Key Laboratory for Particle Physics and Cosmology (SKLPPC), Shanghai Jiao Tong University, 800 Dongchuan Road, Shanghai 200240, China}

\author{Danning Liu}
\affiliation{Tsung-Dao Lee Institute, Shanghai Jiao Tong University, 1 Lisuo Road, Shanghai 201210, China}
\affiliation{Institute of Nuclear and Particle Physics, School of Physics and Astronomy, 800 Dongchuan Road, Shanghai 200240, China}
\affiliation{Key Laboratory for Particle Astrophysics and Cosmology (MOE), Shanghai Key Laboratory for Particle Physics and Cosmology (SKLPPC), Shanghai Jiao Tong University, 800 Dongchuan Road, Shanghai 200240, China}

\author{Kun Liu}\thanks{These authors contributed equally to this work.}
\affiliation{Tsung-Dao Lee Institute, Shanghai Jiao Tong University, 1 Lisuo Road, Shanghai 201210, China}
\affiliation{Institute of Nuclear and Particle Physics, School of Physics and Astronomy, 800 Dongchuan Road, Shanghai 200240, China}
\affiliation{Key Laboratory for Particle Astrophysics and Cosmology (MOE), Shanghai Key Laboratory for Particle Physics and Cosmology (SKLPPC), Shanghai Jiao Tong University, 800 Dongchuan Road, Shanghai 200240, China}

\author{Siyuan Song}
\affiliation{Institute of Nuclear and Particle Physics, School of Physics and Astronomy, 800 Dongchuan Road, Shanghai 200240, China}
\affiliation{Key Laboratory for Particle Astrophysics and Cosmology (MOE), Shanghai Key Laboratory for Particle Physics and Cosmology (SKLPPC), Shanghai Jiao Tong University, 800 Dongchuan Road, Shanghai 200240, China}
\affiliation{Tsung-Dao Lee Institute, Shanghai Jiao Tong University, 1 Lisuo Road, Shanghai 201210, China}

\author{Tong Sun}
\affiliation{Tsung-Dao Lee Institute, Shanghai Jiao Tong University, 1 Lisuo Road, Shanghai 201210, China}
\affiliation{Institute of Nuclear and Particle Physics, School of Physics and Astronomy, 800 Dongchuan Road, Shanghai 200240, China}
\affiliation{Key Laboratory for Particle Astrophysics and Cosmology (MOE), Shanghai Key Laboratory for Particle Physics and Cosmology (SKLPPC), Shanghai Jiao Tong University, 800 Dongchuan Road, Shanghai 200240, China}

\author{Jiannan Tang}
\affiliation{Institute of Nuclear and Particle Physics, School of Physics and Astronomy, 800 Dongchuan Road, Shanghai 200240, China}
\affiliation{Key Laboratory for Particle Astrophysics and Cosmology (MOE), Shanghai Key Laboratory for Particle Physics and Cosmology (SKLPPC), Shanghai Jiao Tong University, 800 Dongchuan Road, Shanghai 200240, China}

\author{Yufeng Wang}
\affiliation{Tsung-Dao Lee Institute, Shanghai Jiao Tong University, 1 Lisuo Road, Shanghai 201210, China}
\affiliation{Institute of Nuclear and Particle Physics, School of Physics and Astronomy, 800 Dongchuan Road, Shanghai 200240, China}
\affiliation{Key Laboratory for Particle Astrophysics and Cosmology (MOE), Shanghai Key Laboratory for Particle Physics and Cosmology (SKLPPC), Shanghai Jiao Tong University, 800 Dongchuan Road, Shanghai 200240, China}

\author{Zhen Wang}
\affiliation{Tsung-Dao Lee Institute, Shanghai Jiao Tong University, 1 Lisuo Road, Shanghai 201210, China}
\affiliation{Institute of Nuclear and Particle Physics, School of Physics and Astronomy, 800 Dongchuan Road, Shanghai 200240, China}
\affiliation{Key Laboratory for Particle Astrophysics and Cosmology (MOE), Shanghai Key Laboratory for Particle Physics and Cosmology (SKLPPC), Shanghai Jiao Tong University, 800 Dongchuan Road, Shanghai 200240, China}

\author{Weihao Wu}
\affiliation{Institute of Nuclear and Particle Physics, School of Physics and Astronomy, 800 Dongchuan Road, Shanghai 200240, China}
\affiliation{Key Laboratory for Particle Astrophysics and Cosmology (MOE), Shanghai Key Laboratory for Particle Physics and Cosmology (SKLPPC), Shanghai Jiao Tong University, 800 Dongchuan Road, Shanghai 200240, China}

\author{Haijun Yang}
% \email[Corresponding author, ]{Haijun Yang, haijun.yang@sjtu.edu.cn}{}
\affiliation{Institute of Nuclear and Particle Physics, School of Physics and Astronomy, 800 Dongchuan Road, Shanghai 200240, China}
\affiliation{Key Laboratory for Particle Astrophysics and Cosmology (MOE), Shanghai Key Laboratory for Particle Physics and Cosmology (SKLPPC), Shanghai Jiao Tong University, 800 Dongchuan Road, Shanghai 200240, China}
\affiliation{Tsung-Dao Lee Institute, Shanghai Jiao Tong University, 1 Lisuo Road, Shanghai 201210, China}

\author{Yuming Lin}
\affiliation{Tsung-Dao Lee Institute, Shanghai Jiao Tong University, 1 Lisuo Road, Shanghai 201210, China}
\affiliation{Institute of Nuclear and Particle Physics, School of Physics and Astronomy, 800 Dongchuan Road, Shanghai 200240, China}
\affiliation{Key Laboratory for Particle Astrophysics and Cosmology (MOE), Shanghai Key Laboratory for Particle Physics and Cosmology (SKLPPC), Shanghai Jiao Tong University, 800 Dongchuan Road, Shanghai 200240, China}

\author{Rui Yuan}
\affiliation{Tsung-Dao Lee Institute, Shanghai Jiao Tong University, 1 Lisuo Road, Shanghai 201210, China}
\affiliation{Institute of Nuclear and Particle Physics, School of Physics and Astronomy, 800 Dongchuan Road, Shanghai 200240, China}
\affiliation{Key Laboratory for Particle Astrophysics and Cosmology (MOE), Shanghai Key Laboratory for Particle Physics and Cosmology (SKLPPC), Shanghai Jiao Tong University, 800 Dongchuan Road, Shanghai 200240, China}

\author{Yulei Zhang}
\affiliation{Institute of Nuclear and Particle Physics, School of Physics and Astronomy, 800 Dongchuan Road, Shanghai 200240, China}
\affiliation{Key Laboratory for Particle Astrophysics and Cosmology (MOE), Shanghai Key Laboratory for Particle Physics and Cosmology (SKLPPC), Shanghai Jiao Tong University, 800 Dongchuan Road, Shanghai 200240, China}

\author{Yunlong Zhang}
\affiliation{State Key Laboratory of Particle Detection and Electronics, University of Science and Technology of China, Hefei 230026, China}
\affiliation{Department of Modern Physics, University of Science and Technology of China, Hefei 230026, China}

\author{Baihong Zhou}
\affiliation{Tsung-Dao Lee Institute, Shanghai Jiao Tong University, 1 Lisuo Road, Shanghai 201210, China}
\affiliation{Institute of Nuclear and Particle Physics, School of Physics and Astronomy, 800 Dongchuan Road, Shanghai 200240, China}
\affiliation{Key Laboratory for Particle Astrophysics and Cosmology (MOE), Shanghai Key Laboratory for Particle Physics and Cosmology (SKLPPC), Shanghai Jiao Tong University, 800 Dongchuan Road, Shanghai 200240, China}

\author{Xuliang Zhu}
\affiliation{Tsung-Dao Lee Institute, Shanghai Jiao Tong University, 1 Lisuo Road, Shanghai 201210, China}
\affiliation{Institute of Nuclear and Particle Physics, School of Physics and Astronomy, 800 Dongchuan Road, Shanghai 200240, China}
\affiliation{Key Laboratory for Particle Astrophysics and Cosmology (MOE), Shanghai Key Laboratory for Particle Physics and Cosmology (SKLPPC), Shanghai Jiao Tong University, 800 Dongchuan Road, Shanghai 200240, China}

\author{Yifan Zhu}
\affiliation{Institute of Nuclear and Particle Physics, School of Physics and Astronomy, 800 Dongchuan Road, Shanghai 200240, China}
\affiliation{Key Laboratory for Particle Astrophysics and Cosmology (MOE), Shanghai Key Laboratory for Particle Physics and Cosmology (SKLPPC), Shanghai Jiao Tong University, 800 Dongchuan Road, Shanghai 200240, China}
\affiliation{Tsung-Dao Lee Institute, Shanghai Jiao Tong University, 1 Lisuo Road, Shanghai 201210, China}

\begin{abstract}
This paper presents the design and optimization of a LYSO crystal electromagnetic calorimeter (ECAL) for the DarkSHINE experiment, which aims to search for dark photons as potential mediators of dark forces. The ECAL design was evaluated through comprehensive simulations, focusing on optimizing dimensions, material selection, energy distribution, and energy resolution. The ECAL configuration consists of 21$\times$21$\times$11 LYSO crystals, each measuring 2.5$\times$2.5$\times$4 cm$^3$, arranged in a staggered layout to improve signal detection efficiency. A 4 GeV energy dynamic range was established to ensure accurate energy measurements without saturation, which is essential for background rejection and signal identification. A detailed digitization model was developed to simulate the scintillation, SiPM, and ADC behaviors, providing a more realistic representation of detector’s performance. Additionally, the study assessed radiation damage in the ECAL region, highlighting the necessity of radiation-resistant scintillators and silicon sensors.
\end{abstract}


\maketitle
\flushbottom

\section{Introduction}
\label{sec:intro}

Dark matter (DM) \cite{Hooper:2007kb, Clowe:2006} remains one of the most compelling mysteries in cosmology and particle physics, evident from its gravitational effects on visible matter and the cosmic microwave background. Theoretical models suggest that DM was produced through thermal processes in the early universe, with the "freeze-out" mechanism\cite{Du:2021jcj} explaining its current observed density, positing a probable mass range from a few MeV to several TeV. Despite extensive searches, the specific properties and particle nature of DM remain elusive. Experiments such as XENONnT\cite{XENON:2023cxc}, PandaX\cite{PandaX-4T:2021bab}, CDEX\cite{Wang_2021}, LUX-ZEPLIN\cite{PhysRevLett.131.041002}, AMS\cite{Giovacchini:2020vxz}, DAMPE\cite{Kyratzis:2022gvg} and LHC\cite{Giagu:2019fmp}  have narrowed the parameter space for weakly interacting massive particles (WIMPs)\cite{Griest:1989wd, Ho:2012ug, Steigman:2013yua, Boehm:2013jpa, Nollett:2013pwa, Nollett:2014lwa, Serpico:2004nm} in the GeV to TeV range, yet these particles remain undetected\cite{Billard:2021uyg}.

Given that high-mass DM candidate particles are yet to be detected, there is increasing interest in the sub-GeV mass range, which presents significant detection challenges due to the minimal interaction cross-sections of such light particles with ordinary matter. This has necessitated the development of innovative detection techniques distinct from traditional DM detection methods. Accelerator-based experiments are particularly promising, leveraging high-energy particle collisions to produce and detect dark photons, a hypothetical mediator between visible and dark matter\cite{Holdom:1985ag, Foot:1991kb, Fuyuto:2019vfe, Choi:2020pyy, Cheng_2022}. Facilities such as CERN's NA64\cite{Banerjee:2019pds}, LHC\cite{2024138378}, BELLE-II\cite{PhysRevLett.130.071804}, BES-III\cite{PhysRevD.100.115016, Prasad:2019ris}, and the proposed LDMX\cite{LDMX:2018cma, LDMX:2019gvz, Berlin:2018bsc} experiment aim to explore these possibilities.

\begin{figure}[htbp]
\centering  %图片全局居中
\subfigure[]{
\includegraphics[width=0.2\textwidth]{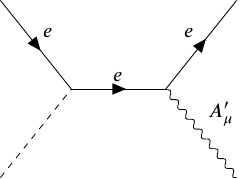}}
\subfigure[]{
\includegraphics[width=0.2\textwidth]{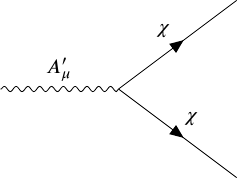}}
\caption{\label{fig:DP}~(a) Production of dark photons via bremsstrahlung. (b) Decay of dark photons into "invisible" modes, where the dark photon decays into dark matter particles. \cite{Fabbrichesi_2021}}
\end{figure}

The DarkSHINE experiment\cite{Chen:2022liu, SLi}, designed to operate under the minimal dark photon model, proposes a novel approach to detecting these elusive particles. Figure~\ref{fig:DP} illustrates the conceptual mechanism of dark photon production through dark bremsstrahlung and their subsequent decay into invisible dark matter, forming the experimental basis for DarkSHINE. This experiment leverages the high-repetition-rate single-electron beam provided by the Shanghai High Repetition-Rate XFEL and Extreme Light Facility (SHINE)\cite{Wan:2022het, Zhao:2017ood}, optimized for detecting subtle signals that hint the presence of dark photons.

This paper delineates the design principles of the Electromagnetic Calorimeter (ECAL) for DarkSHINE experiment, focusing on the selection of materials and structural layout. Through simulation studies, we optimized the ECAL volume, achieving a balance between signal efficiency and cost-effectiveness. We also investigated the dynamic range of energy measurement in the ECAL channels. Insights from these analyses have guided the development of potential triggering strategies for future detector. To ensure the simulated detector response closely mirrors that of a real detector, we implemented a digitization model and studied the ECAL's energy resolution. Furthermore, we evaluated the radiation damage in ECAL region, highlighting the necessity for crystals and silicon sensors that maintain high performance in high-radiation environments.

\section{DarkSHINE experiment and ECAL design}

DarkSHINE is a fixed-target experiment that focuses on the bremsstrahlung production of dark photons and measuring their invisible decay. It utilizes a high-repetition-rate single-electron beam provided by SHINE, which is currently under construction. The beam is expected to achieve an energy of 8GeV and a repetition rate of 10MHz. This corresponds to $3 \times 10^{14}$ electrons-on-target events during one year of the DarkSHINE experiment's commissioning.

\begin{figure}[htbp]
\centering
\includegraphics[width=1\linewidth]{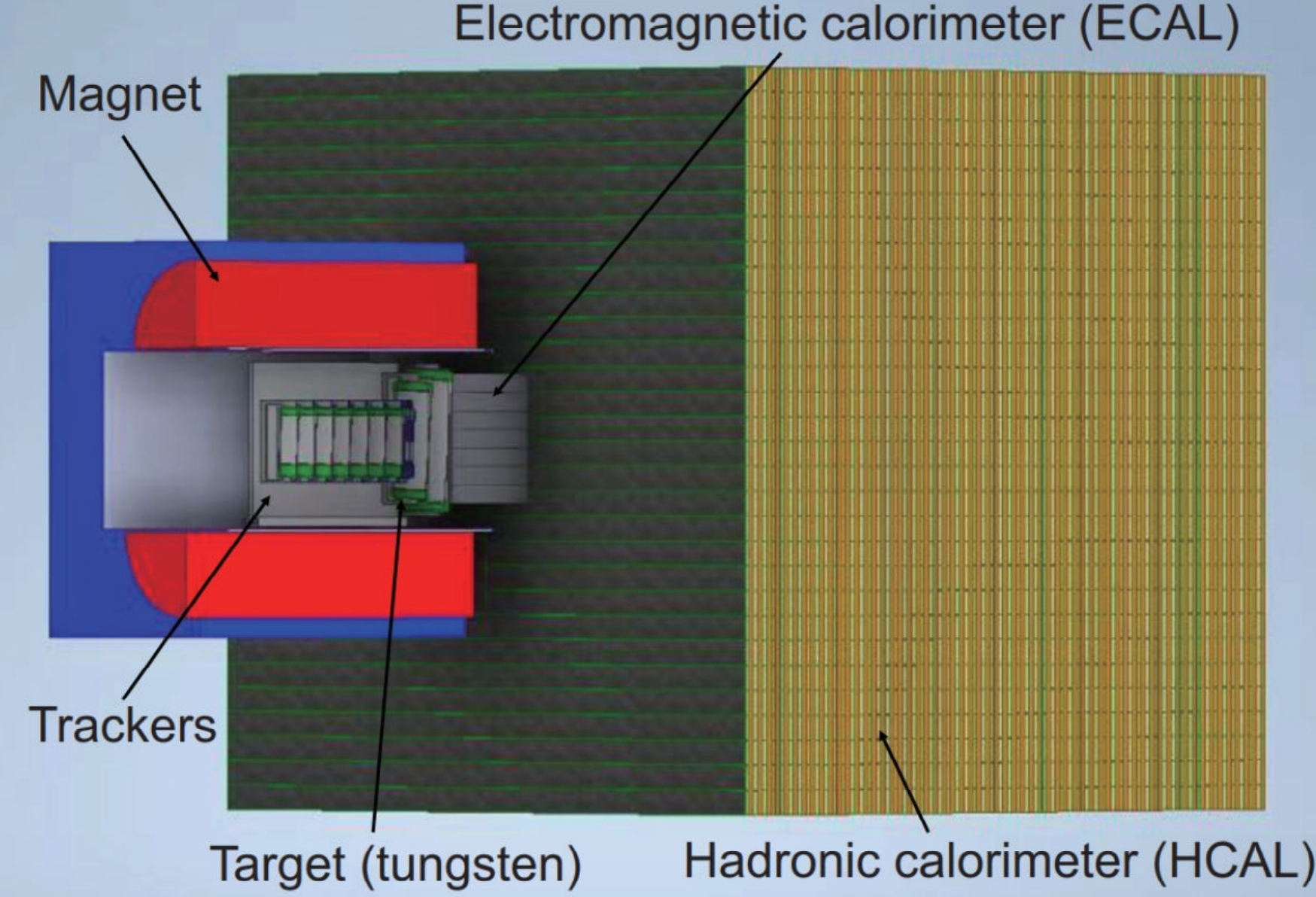}
\caption{\label{fig:DarkSHINEDetector}Schematic of the DarkSHINE detector system, illustrating the primary components used for dark photon detection. The beam enters the center of the detector along the direction from left to right in the diagram. The setup includes the magnet (red) and its supportor (blue), a tagging tracker with seven layers (equal-sized tiles with gray in the middle and green at both ends) inside the magnet, a recoil tracker with six layers (unequal-sized tiles) behind the tagging tracker, a tungsten target sandwiched between the tagging tracker and the recoil tracker, an electromagnetic calorimeter (small gray cube) behind the recoil tracker, and a hadronic calorimeter (big yellow cube) behind the electromagnetic calorimeter. \cite{Chen:2022liu}}
\end{figure}

The primary challenge for DarkSHINE is to use information from various detectors to minimize background contributions while preserving the dark photon signal. Figure~\ref{fig:DarkSHINEDetector} shows the detector system of DarkSHINE. It primarily consists of a tungsten target and three sub-detectors: a tracker, an electromagnetic calorimeter (ECAL), and a hadronic calorimeter (HCAL). Additionally, a single electron beam with a spatial distribution radius of 3 cm and a maximum non-uniform magnetic field of 1.5 Tesla in the tracker region were used in the simulations. The tracker measures the momentum and trajectory of electrons, consisting of a total of thirteen layers of silicon strips. Seven layers are placed in front of the target to tag the incident electron (tagging tracker), and six layers are placed behind the target (recoil tracker). The tungsten target (0.1 X$_0$) is sandwiched between the tagging tracker and the recoil tracker. A crystal ECAL positioned behind the recoil tracker measures the deposited energy of recoil electrons and photons. The recoil electrons are defined as the state of incident electrons after interacting with or passing through the target. Then, a scintillator-steel based sampling HCAL is placed behind the ECAL to capture and veto the backgrounds, in particular the neutral hadrons and muons.

To measure the invisible decay of dark photons, it is essential to precisely measure the energy of recoil electrons after each collision to determine if there is a significant energy loss. The core detector of DarkSHINE, the ECAL designed for DarkSHINE, is a homogeneous LYSO crystal calorimeter. LYSO\cite{1239590, Butler:2019rpu, Kalinnikov:2023coj} is an excellent choice for our crystal material, not only because of its high light yield, which is critical for electromagnetic resolution, but also due to its rapid scintillation decay time (40 ns), which is vital for handling the extremely high event rate we face. Furthermore, the ECAL, especially its central region, is exposed to a considerable radiation dose, necessitating materials with good radiation resistance. LYSO has shown exceptional durability in such conditions, establishing it as an optimal selection for DarkSHINE ECAL applications.

\begin{figure}[htbp]
\centering
\includegraphics[width=0.8\linewidth]{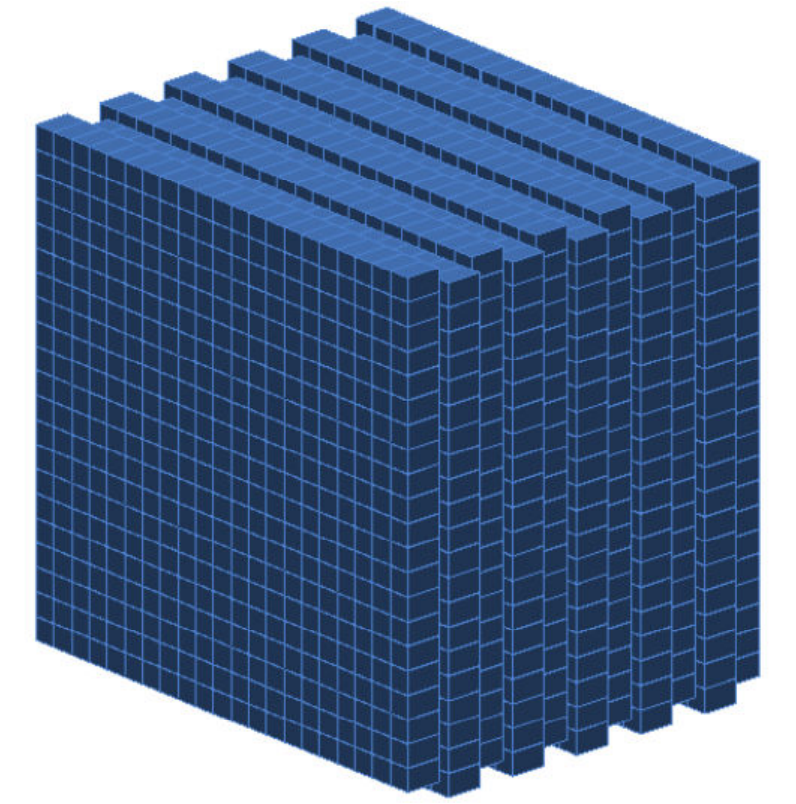}
    \caption{\label{fig:ECAL}~Configuration of the Electromagnetic Calorimeter (ECAL) in DarkSHINE. The ECAL consists of an array of LYSO crystals, each coupled with one SiPM for readout at the end. The schematic shows the segmentation and staggered layout of the crystals within the ECAL.}
\end{figure}

The structure of ECAL is illustrated in Figure~\ref{fig:ECAL}. The transverse dimension of the ECAL is about 52.5$\times$52.5 cm$^2$, with a dimension along the beam direction around 44 cm, approximately equivalent to 39 radiation lengths. The substantial depth ensures excellent energy containment in the ECAL, guaranteeing that it can absorb nearly all electromagnetic showers, preventing them from leaking into the HCAL. This is crucial because such leakage would result in vetoing dark photon signals by the HCAL. The volume of ECAL is determined by balancing the signal efficiency and crystal cost (Section~\ref{sec:VolOpt}). The ECAL is fully segmented to gather comprehensive information, enhancing its capability to discriminate dark photons. It is composed of 21$\times$21$\times$11 LYSO crystals, with each crystal measuring 2.5$\times$2.5$\times$4 cm$^3$ covered by ESR and read out by SiPMs\cite{KLANNER201936, SIMON201985}. In each layer, the crystals are positioned in a uniform 21$\times$21 square pattern. To enhance detection efficacy and prevent particles from traversing the gaps, the placement of crystals in successive layers is shifted by half the transverse dimension of a crystal. This staggered structure aids in detecting recoil electron with energy loss coming along with dark photon signal. \cite{Yu:2024quo, Xu:2024btf}

\section{Signal and background}
\subsection{Software setup}

The Monte Carlo simulation of the ECAL is performed using the DarkSHINE Software (DSS) framework based on Geant4\cite{GEANT4:2002zbu, Zhu:2023}. This simulation includes the entire detector setup, encompassing the crystal cubes, wrappers, silicon sensors, and support structures, alongside other components like the Tracker and HCAL. This comprehensive setup provides the necessary reference for ECAL reconstruction and the initial analysis cuts to define the appropriate phase space for ECAL analysis. Within DSS, we can simulate all Standard Model background processes involving electron-target interactions and various forms of dark photon processes, as well as their responses in the detector.

The target is used in a "full simulation" scenario where an 8 GeV electron beam strikes the target as proposed in the experiment, and secondary particles traverse all detector components. In particular, an "ECAL-unit simulation" focuses on a single ECAL unit to study its precise effects, directly comparable with laboratory test results during R\&D. All types of simulations are conducted with consistent material setups.

\subsection{Signal and background}

\begin{figure}[htbp]
\centering
\includegraphics[width=1\linewidth]{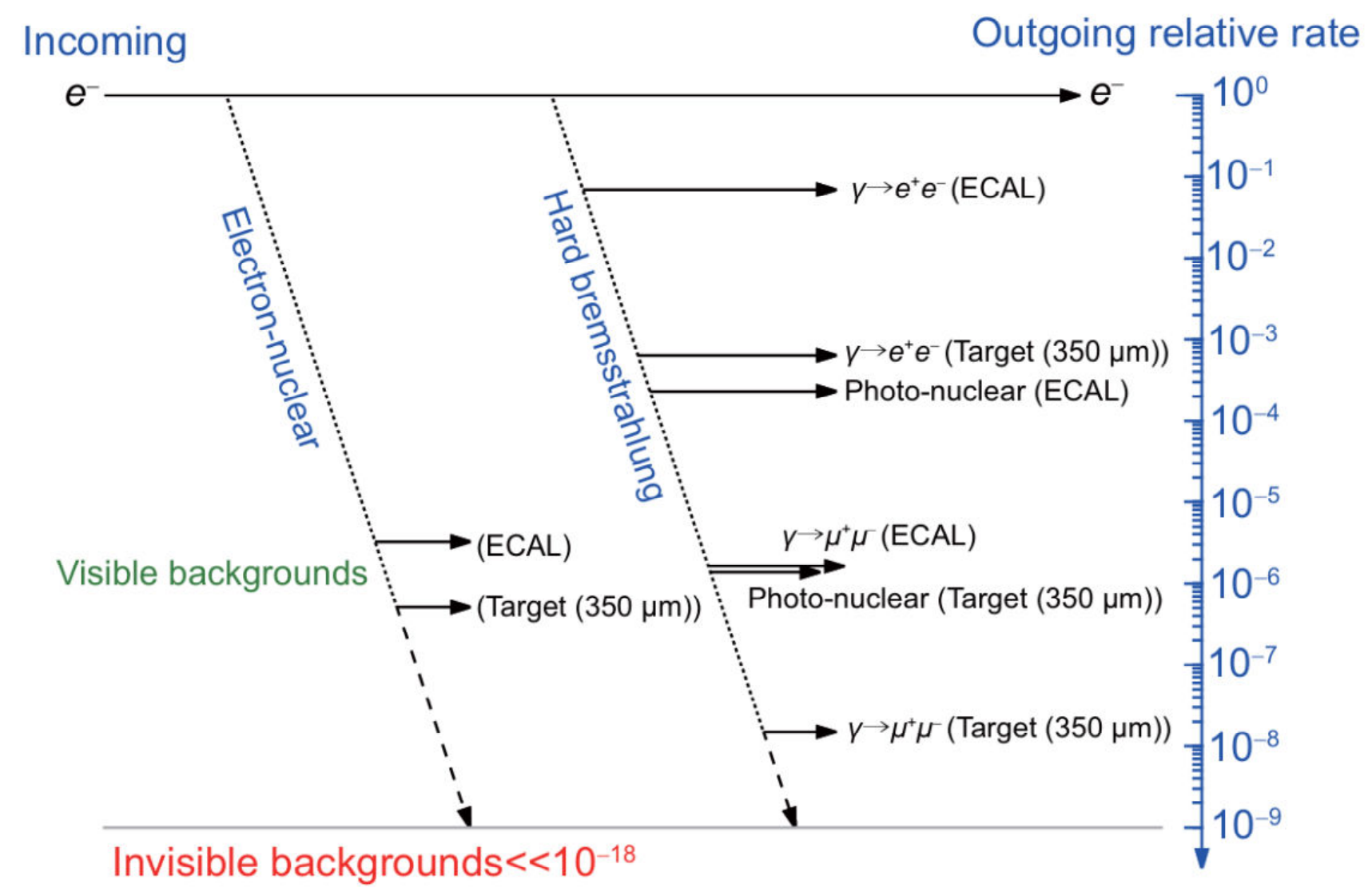}
\caption{\label{fig:BranchingRatio}~Background processes in DarkSHINE experiment, along with their relative rates, under an 8 GeV electron beam. \cite{Chen:2022liu}}
\end{figure}

Figure~\ref{fig:BranchingRatio} illustrates the background processes considered in DarkSHINE experiment, along with their relative rates, under an 8 GeV electron beam. In the majority of cases, the incident electron passes through the target with minimal interacting. Approximately 6.7\% of electrons undergo hard bremsstrahlung, where the bremsstrahlung photons carry away more than 4 GeV from the incident electrons. These bremsstrahlung photons can then interact with the target or ECAL materials via photon-nuclear interactions, occurring with relative rates of 2.31~$\times$~10$^{-4}$ and 1.37~$\times$~10$^{-6}$, respectively. In rare instances, bremsstrahlung photons may convert into a muon pair, with relative rates of 1.63~$\times$~10$^{-6}$ in the ECAL and 1.50~$\times$~10$^{-8}$ in the target. These photons can also produce electron-positron pairs at significantly higher rates. Additionally, electron-nuclear interactions within the ECAL and target materials contribute to the background at relative rates of 3.25~$\times$~10$^{-6}$ and 5.10~$\times$~10$^{-7}$, respectively.~\cite{Chen:2022liu}

\begin{figure}[htbp]
\centering
\includegraphics[width=1\linewidth]{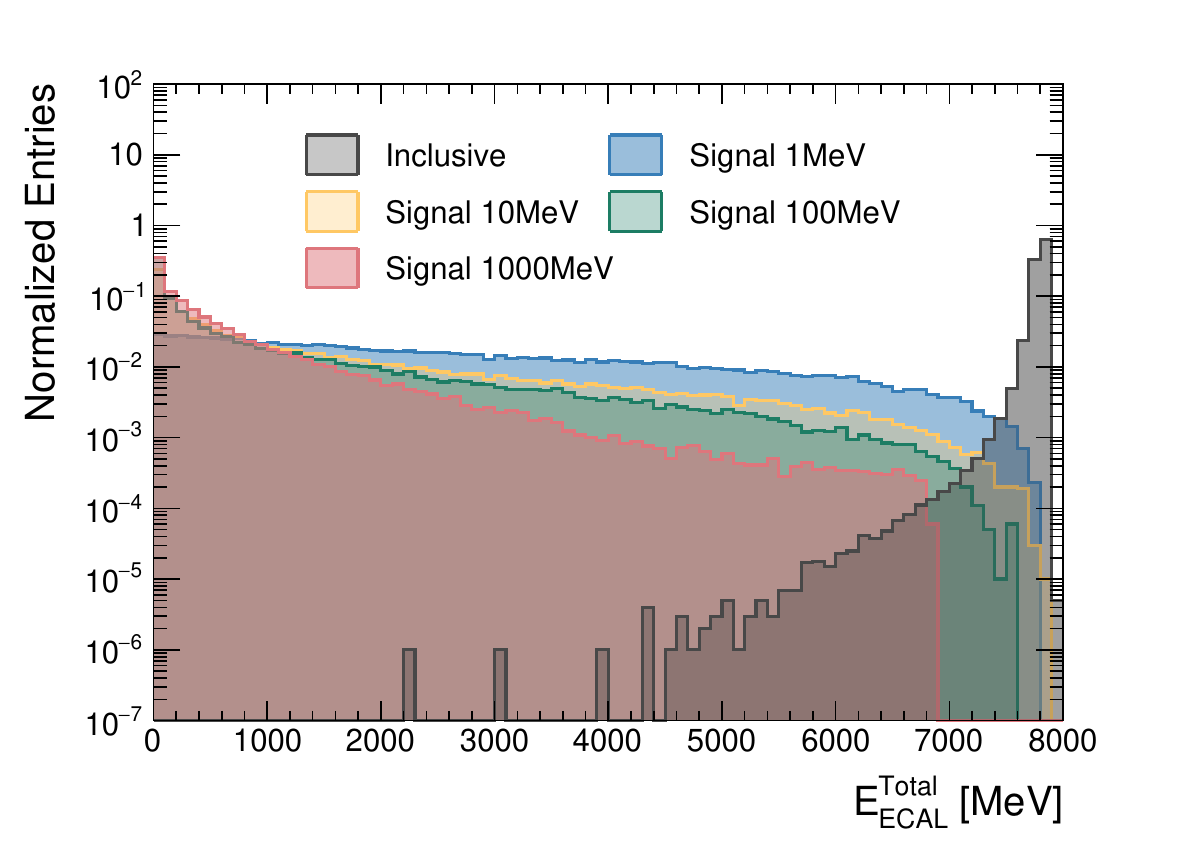}
\caption{\label{fig:Characteristics}~Deposited energy of 8 GeV electrons in ECAL for dark photons and inclusive background processes. In dark photon processes, the energy deposited in ECAL is mostly less than 4 GeV, and it decreases as the mass of the dark photon increases. In contrast, inclusive background process results in significantly higher energy deposition within the ECAL.}
\end{figure}

When a dark photon is produced during the electron-on-target process, most of the incident energy transfers to it, and the recoiled electron deposits its remaining energy in the ECAL. The dark photon decays into dark matter, leaving no signal in the tracker or calorimeter, causing significant energy loss in the ECAL. In Figure~\ref{fig:Characteristics}, more than 75\% of the events have energy deposits in the ECAL that are less than 4 GeV for 1 MeV dark photon signals. As the dark photon mass increases, it carries more energy from the electrons, leading to larger energy loss measured in ECAL. 

The inclusive process encompasses all possible Standard Model physics processes that can occur in the detector, weighted by their respective branching ratios. Compared to the dark photon signal, most events in the inclusive background tend to deposit the majority of their energy in the ECAL. Therefore, the total energy measured by the ECAL becomes a robust criterion for background rejection.

In inclusive process, when the incident electron undergoes hard bremsstrahlung or the produced bremsstrahlung photon converts into an electron-positron pair, these events generate electromagnetic showers in the ECAL, leading to significant energy deposition. However, hard processes involving production of high energy muon pairs or hadrons tend to deposit only a small fraction of energy in the ECAL, but they can be effectively rejected using the tracker and HCAL.

Neutrino-producing background processes, such as Moller scattering ($e^{-}e^{-} \rightarrow e^{-}e^{-}$) followed by a charged-current quasi-elastic (CCQE) reaction ($e^{-}p \rightarrow \nu_{e}n$), neutrino pair production ($e^{-}N \rightarrow e^{-}N\nu\nu$), bremsstrahlung with CCQE, and charge-current exchange with exclusive ($e^{-}p \rightarrow \nu n \pi^{0}$), as mentioned in \cite{Izaguirre:2014bca}, have been found negligible in this experiment due to their low event ratios. Therefore, these backgrounds are currently excluded from the DarkSHINE simulation\cite{Chen:2022liu}.

In summary, the ECAL aims at excluding backgrounds that do not involving hard muons or hadrons.

\section{Volume optimization}
\label{sec:VolOpt}

\begin{figure}[htbp]
\centering
\includegraphics[width=1\linewidth]{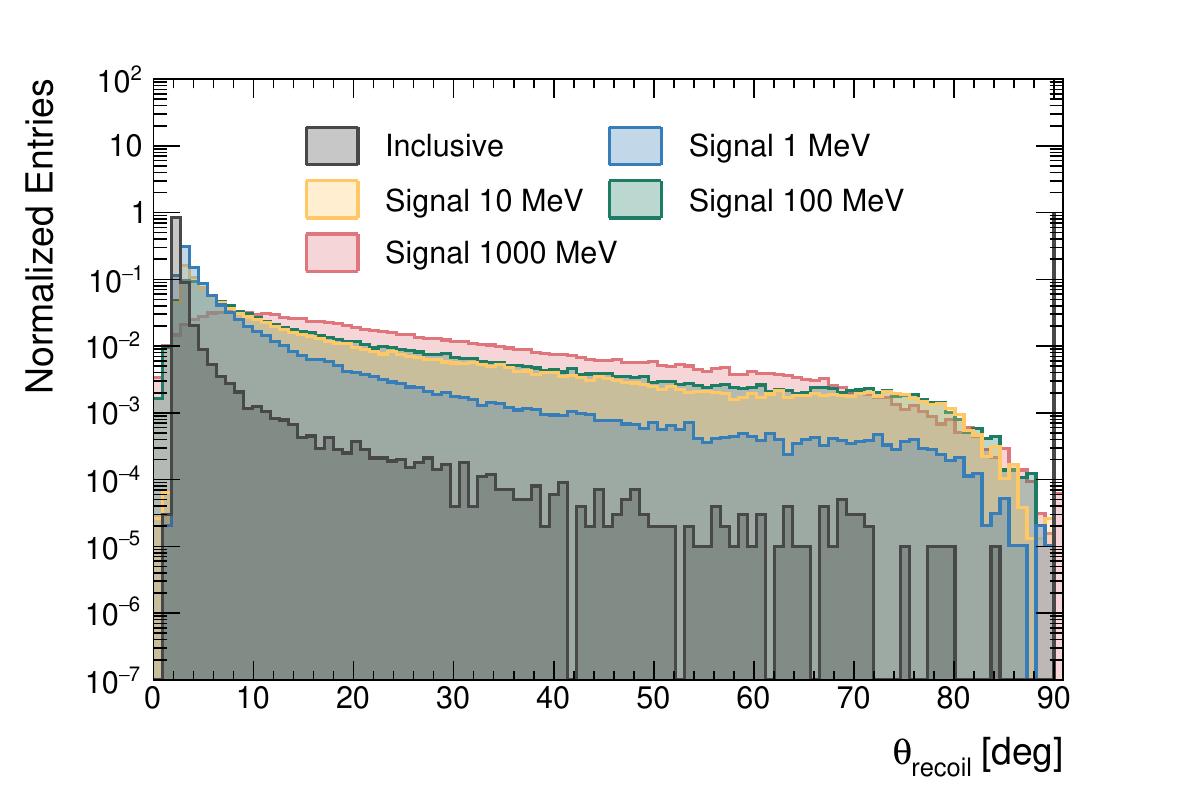}
\caption{\label{fig:RecoilTheta}~The distribution of recoil angles for different processes. Recoil angle is defined as the angle between the particle's momentum direction and the beam direction. In this figure, the ECAL has a cross-sectional area of 52.5$\times$52.5 cm$^2$. Dark photon signals exhibit larger recoil angles compared to inclusive background processes. As the dark photon mass increases, more events tend to impact the peripheral regions of the ECAL.}
\end{figure}

As the mass of dark photons increases, they carry more energy from the incident electron, resulting in a larger recoil angle, which is defined as the angle between the particle momentum direction and the beam direction. Figure~\ref{fig:RecoilTheta} illustrates the distribution of recoil angles for particles striking the front surface of the ECAL for both inclusive events and dark photon processes with varying masses, where the ECAL's cross-sectional area is 52.5$\times$52.5 cm$^2$. The majority of inclusive background processes involve electrons directly traversing the target and striking the central region of the ECAL, despite some divergence in the beam spot and slight displacement of electrons due to the magnetic field. In contrast, signal processes exhibit more dispersed recoil angles. Moreover, with increasing dark photon mass, more events occur in the peripheral region, indicating a higher likelihood of missing the ECAL. But these events can be captured and vetoed by the HCAL. Therefore, increasing the ECAL volume can improve signal efficiency, though cost considerations must also be addressed.

A signal region for the calorimeter is defined in Table~\ref{tab:SignalBox}: the total energy in the ECAL is less than 2.5 GeV, the total energy in the HCAL is less than 30 MeV, and the maximum energy of a single unit in the HCAL is less than 0.1 MeV. This signal region is designed to achieve low background within $3 \times 10^{14}$ electrons-on-target events. It is derived from the combined analysis of all sub-detectors, using thresholds for the ECAL and other sub-detectors, and applying extrapolation methods to exclude all backgrounds\cite{Chen:2022liu, Wang2024}. The optimization results of the ECAL are evaluated by signal efficiency, which is defined as the ratio of the number of events entering the signal region to the number of total simulated events.

\begin{table}[htbp]
\centering
\caption{\label{tab:SignalBox}~Signal region for calorimeters.}
\fontsize{9}{13}\selectfont
\begin{tabular}{cc}
\toprule
\makecell[c]{ECAL} & \makecell[c]{HCAL}\\
\midrule
\makecell[c]{E$_{ECAL}^{total}$ < 2.5GeV} & \makecell[c]{E$_{HCAL}^{total}$ < 30MeV,\\ E$_{HCAL}^{MaxCell}$ < 0.1MeV}\\
\bottomrule
\end{tabular}
\end{table}

\begin{figure}[htbp]
\centering  
\subfigure[]{
\includegraphics[width=0.45\textwidth]{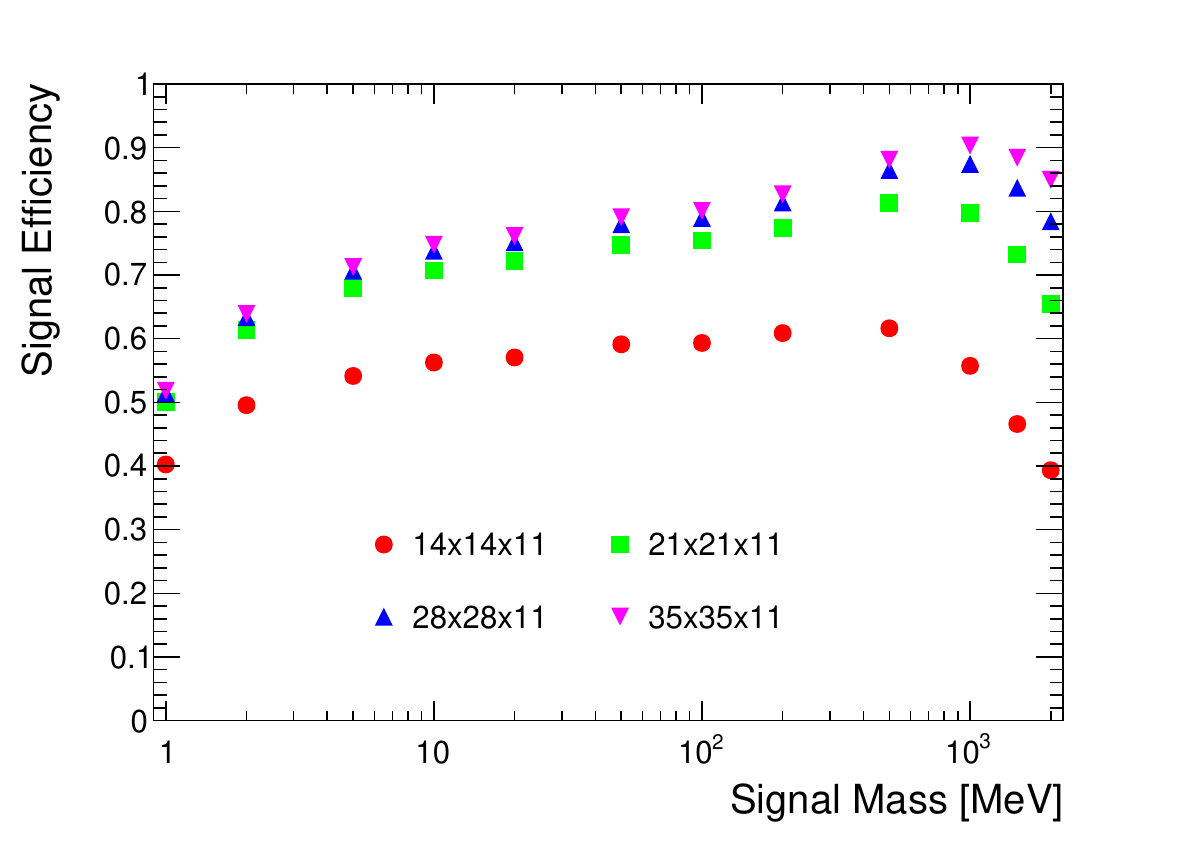}}
\subfigure[]{
\includegraphics[width=0.45\textwidth]{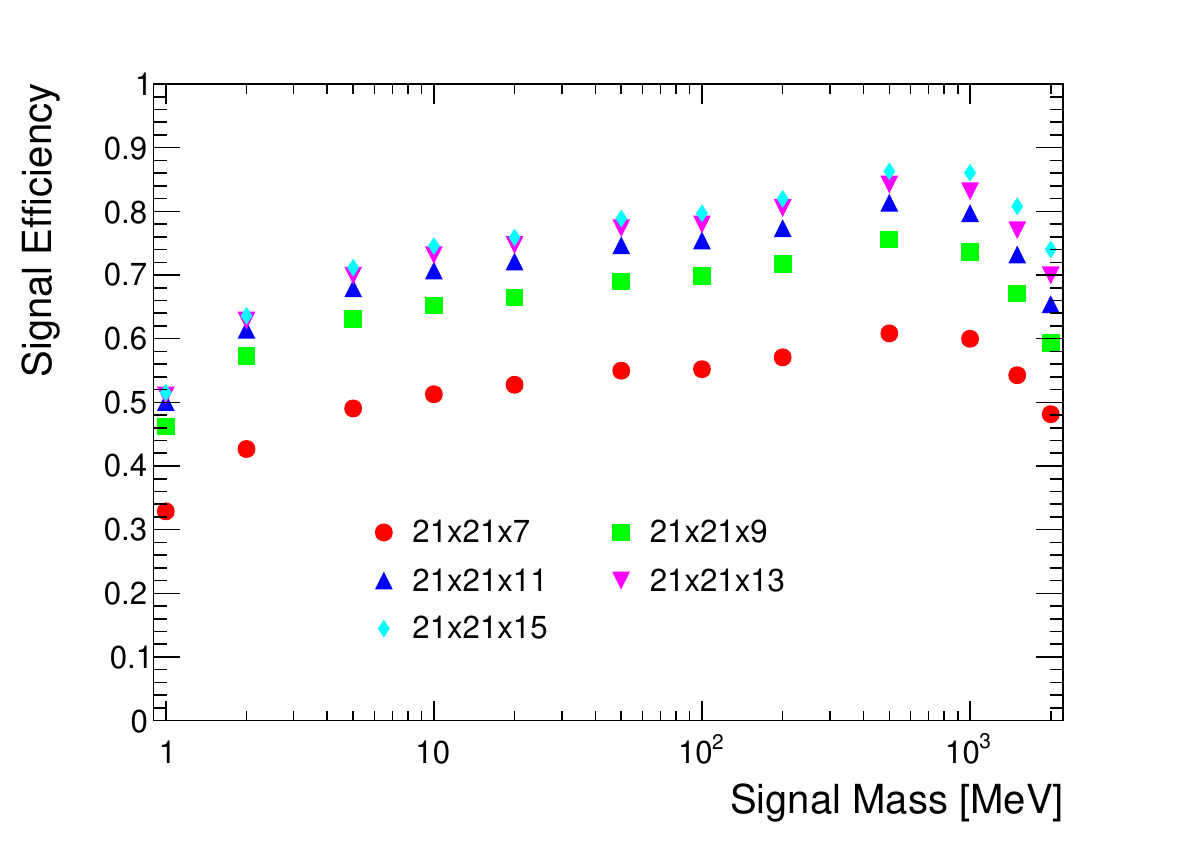}}
\caption{\label{fig:VolumeOpt} Signal efficiency as a function of dark photon mass for various ECAL configurations. The signal efficiency is defined as the ratio of the number of events entering the signal region in Table~\ref{tab:SignalBox}, to the number of total simulated events. In the legends, the format "x$\times$y$\times$z" represents the number of crystals in the ECAL along the transverse-x, transverse-y, and longitudinal-z dimensions, with each crystal measuring 2.5$\times$2.5$\times$4.0 cm$^3$. The beam travels along the longitudinal direction. (a) Optimization of the transverse size of the ECAL, showing signal efficiency for different transverse dimensions. Increasing the ECAL's transverse size can enhance signal efficiency, particularly for higher dark photon masses. (b) Optimization of the longitudinal size of the ECAL, showing signal efficiency for different longitudinal dimensions. It shows that increasing the ECAL's longitudinal size improves signal efficiency.}
\end{figure}

The ECAL size was optimized in two dimensions: transverse dimension and longitudinal dimension, with the longitudinal direction being aligned along the beam direction. During optimization, the individual crystal size remains constant at 2.5$\times$2.5$\times$4.0 cm$^3$, which is determined by the maximum density that our electronics can handle~\cite{guo2024}. 

First, the ECAL's transverse dimensions were adjusted while keeping the longitudinal length constant at 11 layers. Four different calorimeter sizes were simulated. Results in Figure~\ref{fig:VolumeOpt}(a) show that as the transverse size increases, signal efficiency for different dark photon masses also increases. The average signal efficiency for all mass points under each size is listed in Table~\ref{tab:SignalEfficiency_Tran}.

\begin{table}[htbp]
    \centering
    \caption{\label{tab:SignalEfficiency_Tran}Average signal efficiency with varying transverse size of the ECAL.}
    \fontsize{9}{13}\selectfont
    \begin{tabular}{cc}
        \hline
        Number of crystals & Average signal efficiency(\%)\\
        \hline
        14$\times$14$\times$11 & 53.32\\
        21$\times$21$\times$11 & 70.78\\
        28$\times$28$\times$11 & 75.75\\
        35$\times$35$\times$11 & 77.62\\
        \hline
    \end{tabular}
\end{table}

The width of the ECAL was increased by 7 crystal blocks each time. From a design with 14×14×11 crystals to one with 21×21×11 crystals, the ECAL area increased by 2.25 times, resulting in a 17.46\% improvement in signal efficiency. However, when the number of crystals per layer reaches 35×35, the increase in signal efficiency becomes marginal. Considering the cost of crystals, an ECAL with 21×21 crystals is deemed appropriate.

\begin{table}[htbp]
    \centering
    \caption{\label{tab:SignalEfficiency_Long}Average signal efficiency with varying longitudinal size of the ECAL.}
    \fontsize{9}{13}\selectfont
    \begin{tabular}{cc}
        \hline
        Number of crystals & Average signal efficiency(\%)\\
        \hline
        21$\times$21$\times$7  & 52.05\\
        21$\times$21$\times$9  & 66.32\\
        21$\times$21$\times$11 & 71.52\\
        21$\times$21$\times$13 & 74.07\\
        21$\times$21$\times$15 & 76.71\\
        \hline
    \end{tabular}
\end{table}

The impact of the longitudinal size of the detector on signal efficiency was also investigated. Each time the detector volume was changed, the ECAL's transverse dimensions were kept constant as 21×21 crystals, and two layers were added longitudinally. Thus, the number of added crystals remained the same each time. Table~\ref{tab:SignalEfficiency_Long} shows that from the size of 11 layers, the growth in average signal efficiency begins to slow. Therefore, 11 longitudinal layers were chosen as the final size.

Overall, an ECAL size of 52.5$\times$52.5$\times$44 cm$^3$ provides significant signal efficiency while maintaining reasonable cost.

\section{Energy distribution}
\label{sec:EnergyDis}

As a fixed target experiment, incident particles without significant energy loss primarily hit the central area of the ECAL, resulting in substantial energy deposition in the central region's crystals in the first few layers. Additionally, the energy absorbed by crystals in different regions varies greatly, with those closer to the edges typically experiencing minimal energy deposition. To determine the energy dynamic range for the future detector and explore potential triggering methods for the ECAL, the energy deposition in crystals from different regions was investigated.

\subsection{Energy deposition in different regions}

\begin{figure}[htbp]
\centering
\includegraphics[width=1\linewidth]{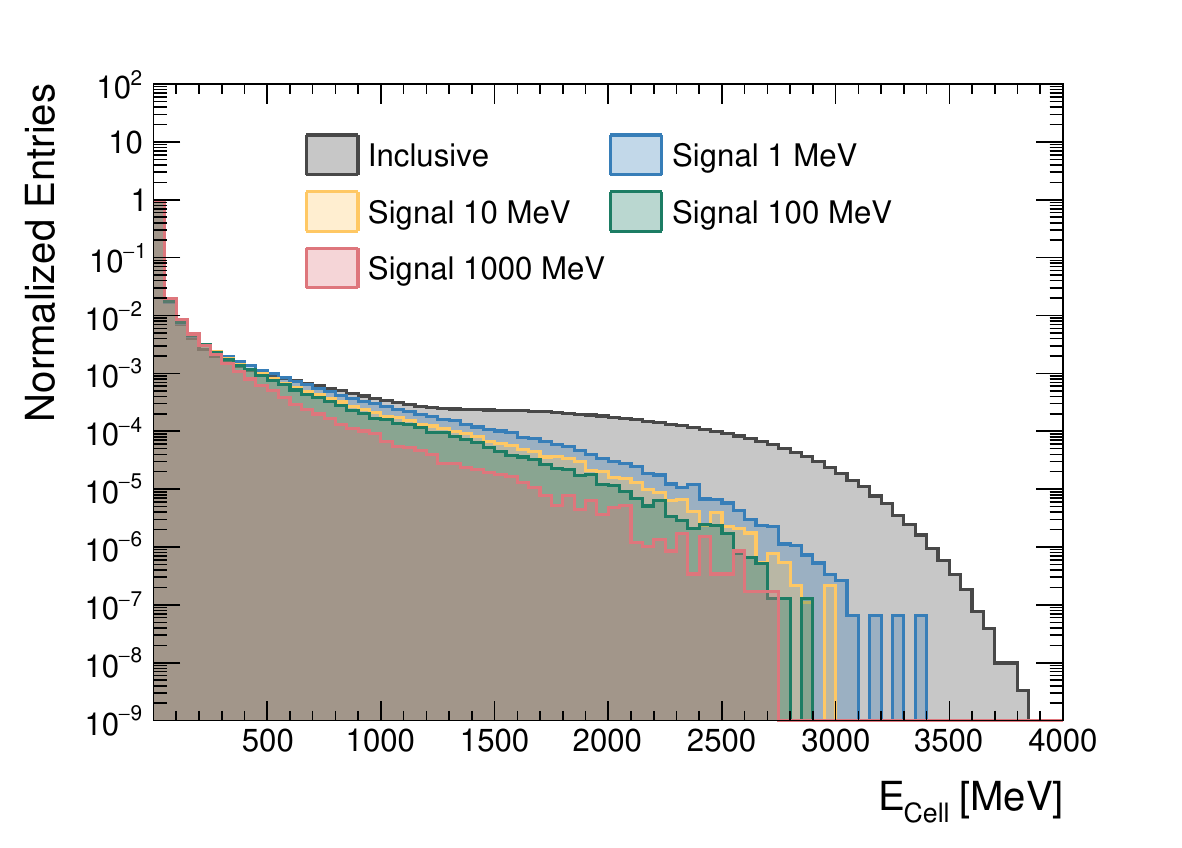}
\caption{\label{fig:ECellTotal}~The energy deposited in individual crystals by an 8 GeV electron beam incident on the ECAL. For inclusive background processes, the energy absorbed by a single crystal can reach up to 4 GeV, while for dark photon signal, it can reach up to 3.5 GeV.}
\end{figure}

With the optimized ECAL volume described in \ref{sec:VolOpt}, the energy deposition in crystals for dark photon signals and background processes is shown in Figure~\ref{fig:ECellTotal}. Inclusive background processes typically lead to larger energy deposits in the crystals, reaching up to 4 GeV. While the energy deposition from dark photon signals is slightly lower, with smaller mass dark photons depositing more energy. A dark photon with a mass of 1 MeV can deposit up to 3.5 GeV in the crystals, as dark photons have a very low probability of carrying minimal kinetic energy, causing the recoiling electron to retain most of the energy.

\begin{figure}[htbp]
\centering
\includegraphics[width=0.8\linewidth]{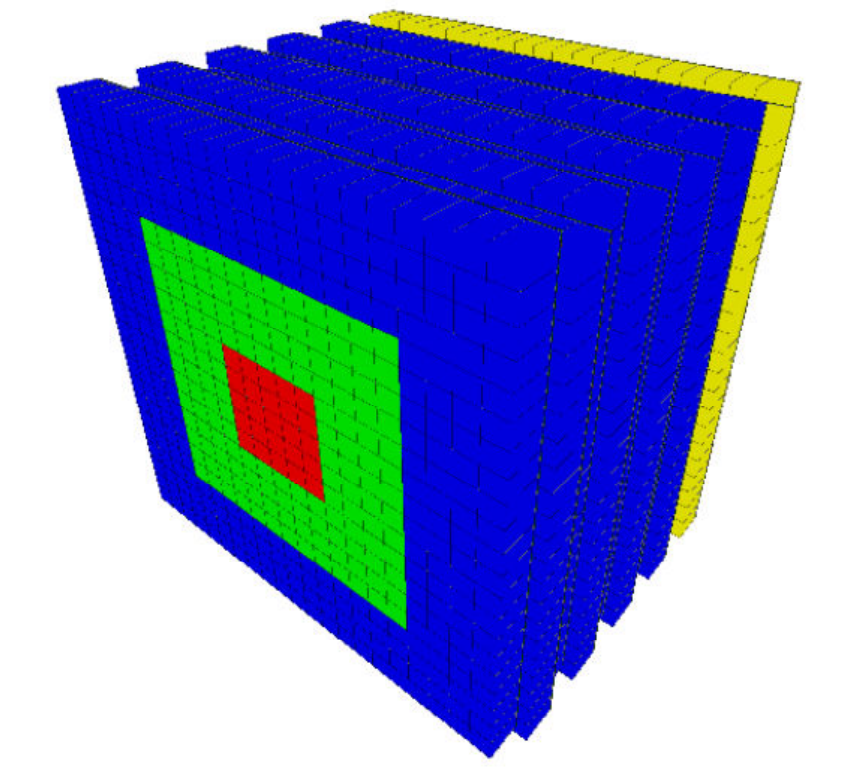}
\caption{\label{fig:geo}~ECAL divided into four regions, from the center to the periphery. Red: region-\uppercase\expandafter{\romannumeral 1}, green: region-\uppercase\expandafter{\romannumeral 2}, blue: region-\uppercase\expandafter{\romannumeral 3}, yellow: region-\uppercase\expandafter{\romannumeral 4}. The beam
enters the center of the detector along the direction from left to right
in the diagram. The beam travels from left to right in the figure, with the beam direction defined as the longitudinal direction (z).}
\end{figure}

\begin{table}[htbp]
    \centering
    \caption{\label{tab:Region}Number of crystals in each region shown in Figure~\ref{fig:geo}}
    \fontsize{9}{13}\selectfont
    \begin{tabular}{ccccc}
        \hline
        Dimension & Region-\uppercase\expandafter{\romannumeral 1} & Region-\uppercase\expandafter{\romannumeral 2} & Region-\uppercase\expandafter{\romannumeral 3} & Region-\uppercase\expandafter{\romannumeral 4}\\
        \hline 
        Transverse-x     & 5  & 13    & 21   & 21\\
        Transverse-y     & 5  & 13    & 21   & 21\\
        Longitudinal-z   & 5  & 7     & 9    & 2 \\
        Total & 125 & 1058 & 2786 & 882 \\
        \hline
    \end{tabular}
\end{table} 

\begin{figure*}[]
    \centering
    \subfigure[Region-\uppercase\expandafter{\romannumeral 1}]{
        \includegraphics[width=0.45\hsize]{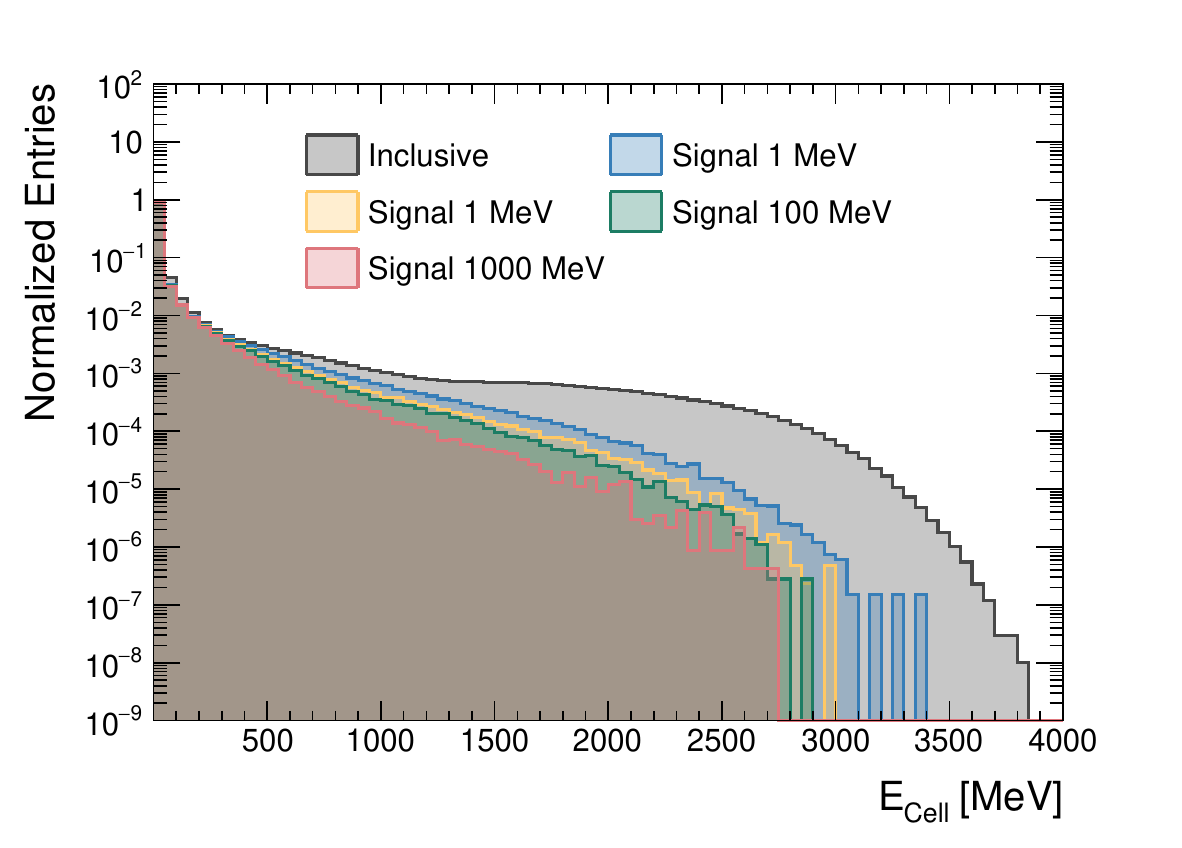}
        \label{fig:first}
    }
    \subfigure[Region-\uppercase\expandafter{\romannumeral 2}]{
        \includegraphics[width=0.45\hsize]{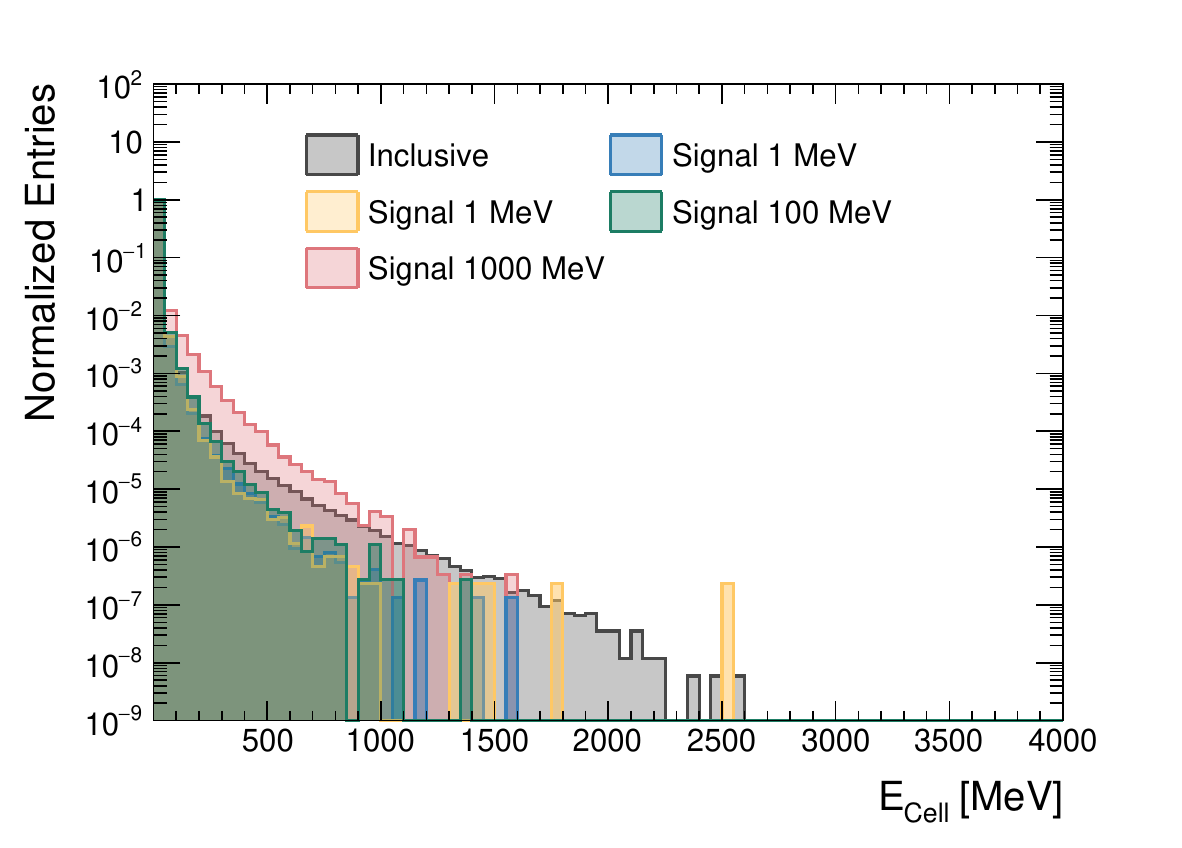}
        \label{fig:second}
    }

    \subfigure[Region-\uppercase\expandafter{\romannumeral 3}]{
        \includegraphics[width=0.45\hsize]{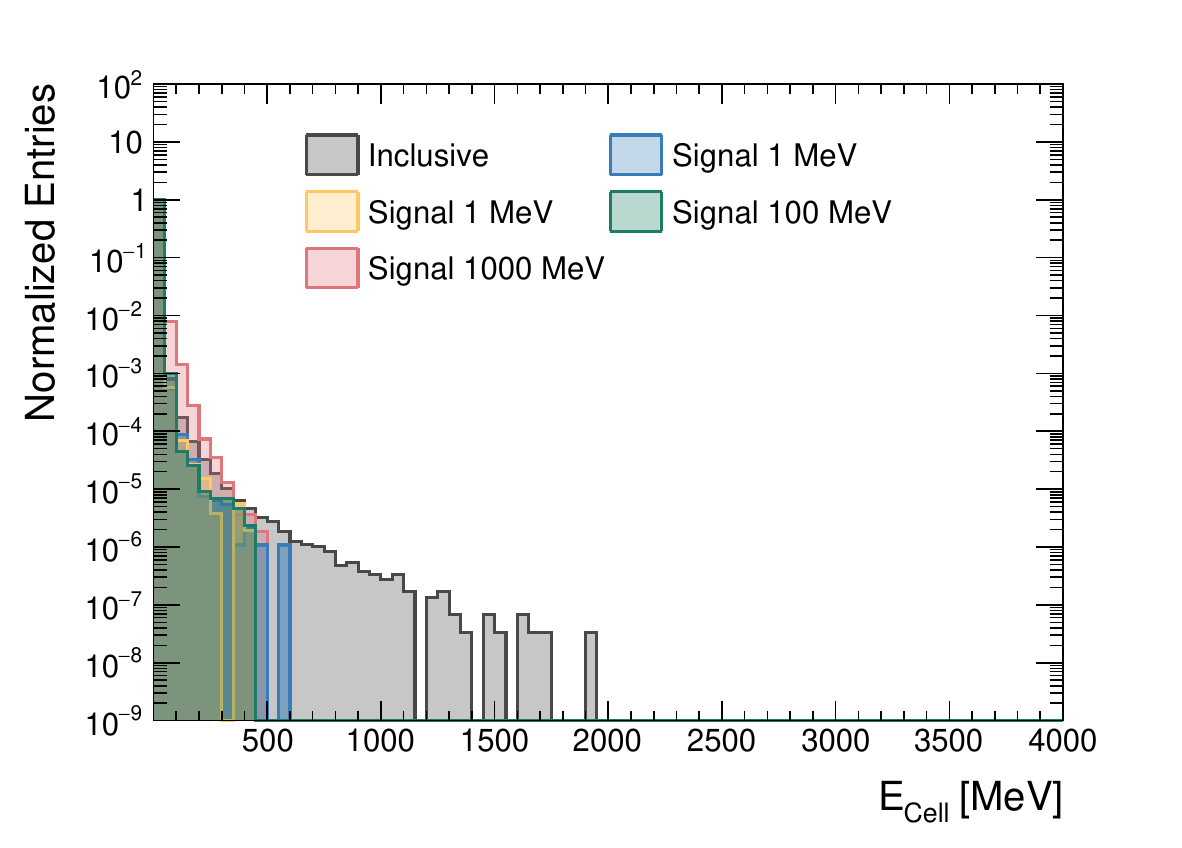}
        \label{fig:third}
    }
    \subfigure[Region-\uppercase\expandafter{\romannumeral 4}]{
        \includegraphics[width=0.45\hsize]{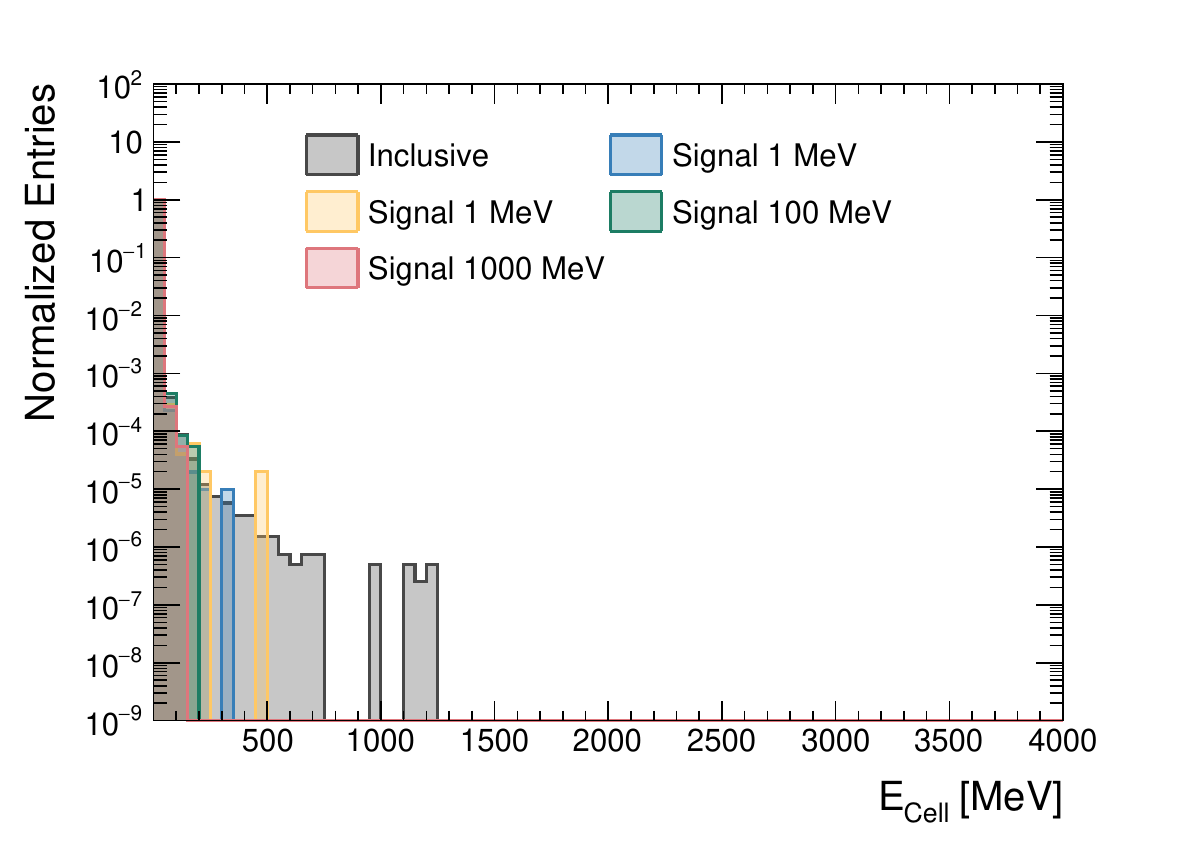}
        \label{fig:fourth}
    }
    \caption{~Energy absorbed by single crystals in the four regions shown in Figure~\ref{fig:geo}. Crystals in the central regions (Region-\uppercase\expandafter{\romannumeral 1} and Region-\uppercase\expandafter{\romannumeral 2}) absorb significantly more energy, while the crystals in Region-\uppercase\expandafter{\romannumeral 3} and Region-\uppercase\expandafter{\romannumeral 4}, farther from the shower center, absorb noticeably less energy.}
    \label{fig:ECellRegion}
\end{figure*}

\begin{figure}[htbp]
\centering
\includegraphics[width=1\linewidth]{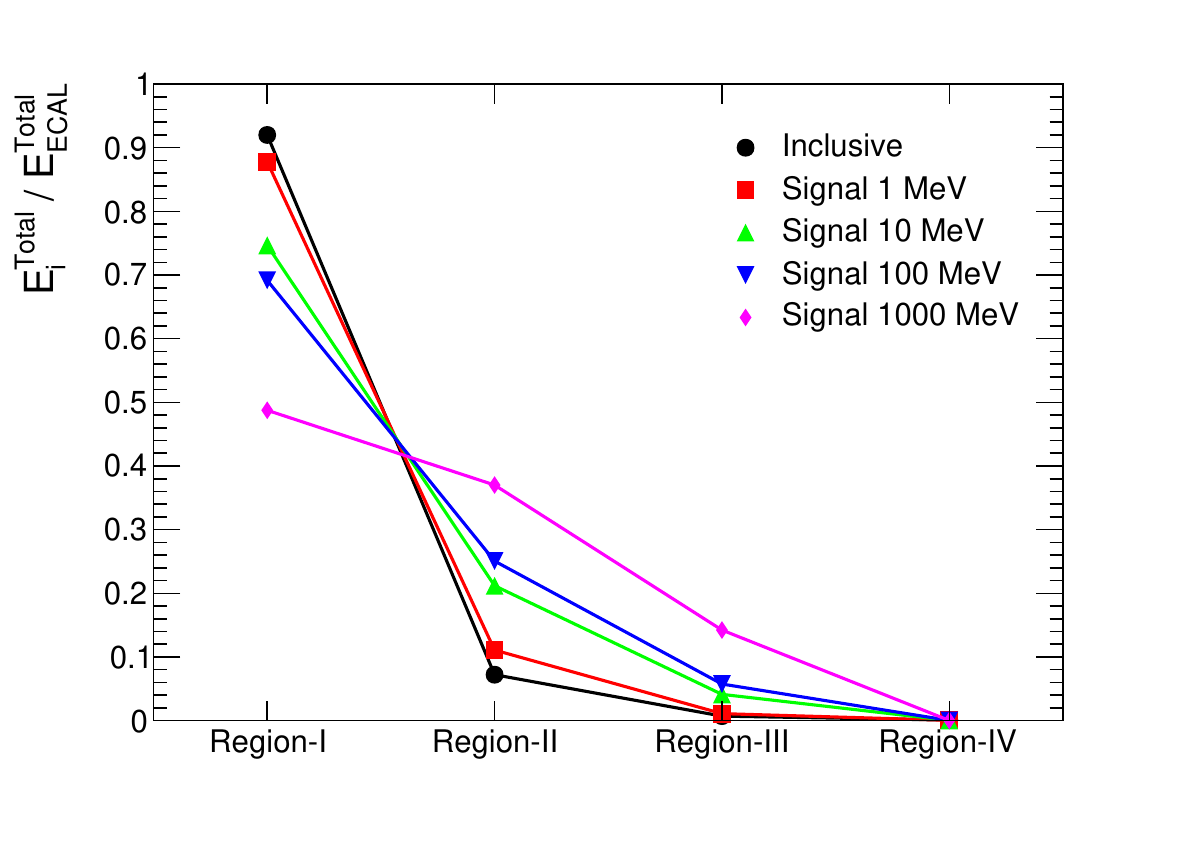}
\caption{\label{fig:ECellRatio}~Energy ratio in different regions shown in Figure~\ref{fig:geo} for various processes. Energy ratio is defined as the ratio of the total energy deposited in the crystals within a specific region ($E_{i}^{Total}$) to the total energy deposited across all crystals in the ECAL ($E_{ECAL}^{Total}$). Both signal and background processes deposit the most energy in Region-\uppercase\expandafter{\romannumeral 1}. However, for dark photon processes, as the dark photon mass increases, a larger proportion of the energy is absorbed in the outer regions.}
\end{figure}

The ECAL was divided into four sections (Figure~\ref{fig:geo}) from the center to the periphery to study energy deposition in different regions. The number of crystals in each region is shown in Table~\ref{tab:Region}. Region-\uppercase\expandafter{\romannumeral 1} (red) is the core area of shower development, containing 125 crystals each with a volume of 2.5$\times$2.5$\times$4 cm$^3$, approximately equivalent to 18 radiation lengths. It covers the entire beam spot with a radius of 3 cm and an additional area of one Molière radius. The energy absorbed by each crystal in these four regions is shown in Figure~\ref{fig:ECellRegion}. Crystals in the central and near-central regions absorb significantly more energy. In contrast, crystals in Region-\uppercase\expandafter{\romannumeral 3} and Region-\uppercase\expandafter{\romannumeral 4}, farther from the shower center, absorb noticeably less energy. For dark photon signals, the energy deposition is even lower than for inclusive background. Figure~\ref{fig:ECellRatio} shows that crystals in Region-\uppercase\expandafter{\romannumeral 1}, with only 2.6\% of the ECAL volume, absorb more than 90\% of the energy for inclusive backgrounds.

\subsection{Energy limits on channels}

The setup of the energy dynamic range directly impacts the accuracy of energy measurements. Generally, a smaller dynamic range allows for higher measurement precision. To identify the minimum dynamic range that satisfies the energy measurement requirements of the ECAL, an energy limit was imposed on each crystal. When the energy deposited in a crystal exceeds this limit, it is capped at the limit, mimicking the saturation behavior observed in a real detector. However, if the energy limit is set too low, it can reduce the total energy measured by the ECAL, potentially causing background events with substantial energy loss to be misidentified as dark photon signals. Since our focus is on events characterized by significant energy loss, we investigated how the energy limit influences the number of events where the total energy measured by the ECAL falls below 4 GeV.

\begin{figure}[htbp]
\centering
\includegraphics[width=1\linewidth]{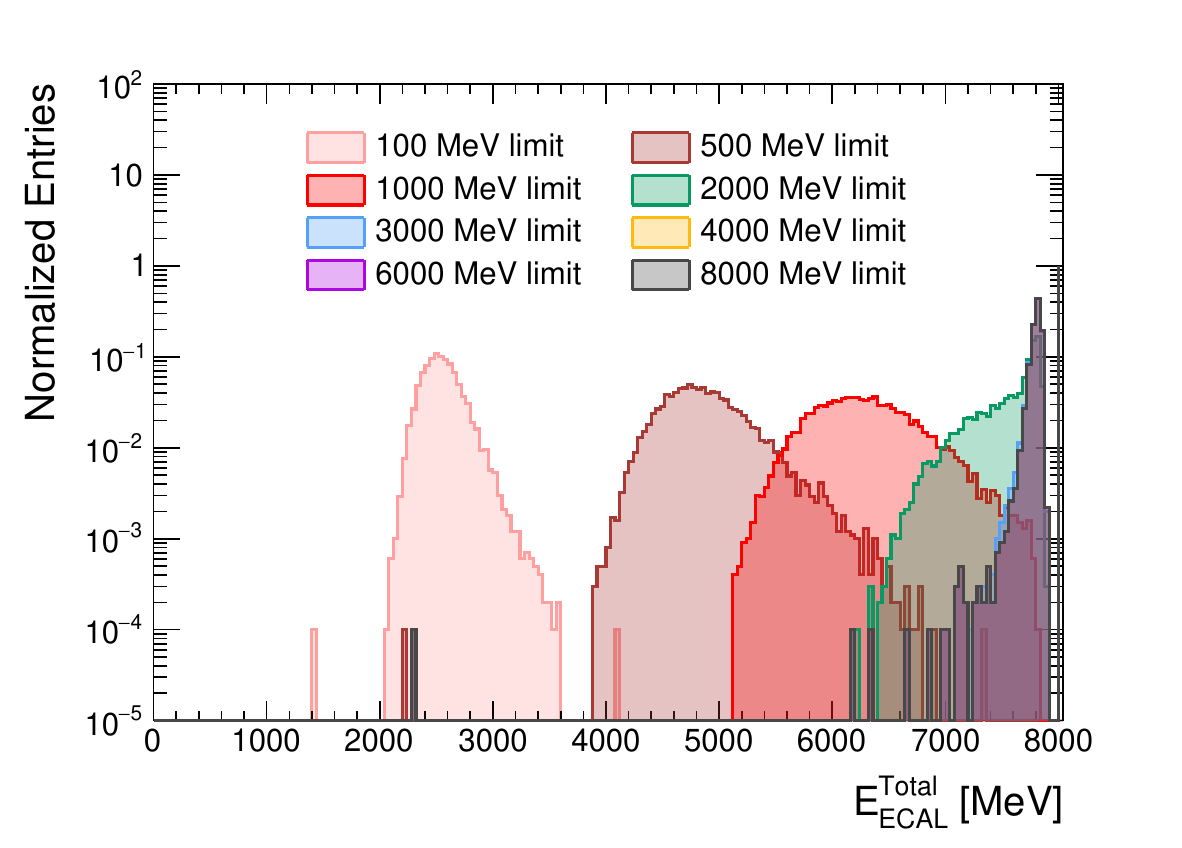}
\caption{\label{fig:ELimit}~ECAL energy distribution of inclusive background after applying different energy limits to channels. The energy limit represents the maximum energy that a single channel can measure, analogous to the saturation in detector electronics. When the limit is set too low, the energy measured by the ECAL can deviate significantly from the incident energy, shifting to a lower energy region. This shift can lead to the misidentification of background events as dark photon signals.}
\end{figure}

Figure~\ref{fig:ELimit} shows the total energy distribution in the ECAL after applying different energy limits to channels. As the energy limit decreases to below 3 GeV, the total energy measured by the ECAL starts to decrease. This is comparable to the saturation effect caused by the insufficient dynamic range of the electronics in detector. If the energy limit is greater than 3 GeV, the energy measured by the ECAL is almost unaffected by the limit. To ensure the accuracy of the ECAL's energy measurement, the dynamic range of individual channels should not impact the total energy. From Figure~\ref{fig:ELimit}, a dynamic range greater than 3 GeV is appropriate. 

\begin{table}[htbp]
\centering
\caption{\label{tab:EnergyLimit}~Ratios between the number of events with ECAL energy less than 4 GeV and the number of total simulated events, calculated after applying various channel energy limits to background and dark photon signals of different masses. As the limit decreases, the energy measured by the ECAL may also decrease, leading to more events falling into the region where the total ECAL energy is less than 4 GeV.}
\fontsize{9}{13}\selectfont
    \begin{minipage}[t]{0.5\textwidth}
    \centering
        \begin{tabular}{cccc}
            \toprule
            \makecell[c]{Energy limit} & \makecell[c]{Inclusive} & \makecell[c]{1 MeV dark photon} & \makecell[c]{10 MeV dark photon}\\
            \midrule
            100 MeV  & 999967/1M & 100\%   & 100\% \\
            500 MeV  & 1066/1M   & 90.57\% & 96.93\% \\
            1000 MeV & 3/1M      & 76.78\% & 91.23\% \\
            2000 MeV & 3/1M      & 76.76\% & 89.89\% \\
            3000 MeV & 3/1M      & 76.76\% & 89.89\% \\
            4000 MeV & 3/1M      & 76.76\% & 89.89\% \\
            6000 MeV & 3/1M      & 76.76\% & 89.89\% \\
            8000 MeV & 3/1M      & 76.76\% & 89.89\% \\
            \bottomrule
        \end{tabular}
    \end{minipage}%
    \hfill
    \begin{minipage}[t]{0.5\textwidth}
    \centering
        \begin{tabular}{ccc}
            \toprule
            \makecell[c]{Energy limit} & \makecell[c]{100 MeV dark photon} & \makecell[c]{1000 MeV dark photon} \\
            \midrule
            100 MeV  & 100\%   & 100\%\\
            500 MeV  & 98.27\% & 99.5\%\\
            1000 MeV & 94.52\% & 98.56\%\\
            2000 MeV & 93.52\% & 98.32\%\\
            3000 MeV & 93.52\% & 98.32\%\\
            4000 MeV & 93.52\% & 98.32\%\\
            6000 MeV & 93.52\% & 98.32\%\\
            8000 MeV & 93.52\% & 98.32\%\\
            \bottomrule
        \end{tabular}
    \end{minipage}
\end{table}

Table~\ref{tab:EnergyLimit} shows the ratios between the number of events with ECAL energy less than 4 GeV and the number of total simulated events. These ratios were derived from background and dark photon signals of varying masses after applying different channel energy limits. As the energy limit decreases, the inclusive background has increasing energy loss, resulting in a growing number of cases where the ECAL energy is less than 4 GeV. The results indicate that a dynamic ranges of 100 MeV or 500 MeV is insufficient, as they demonstrate poor background rejection capability with a statistical sample of one million events. When the energy limit on channels is set to above 500 MeV, nearly all events deposit more than 4 GeV of energy in the ECAL, except for two photon-muon pair processes and one nuclear process, which can be effectively vetoed by the HCAL. This indicates very few crystals exhibit energy depositions exceeding 1 GeV, which is also reflected in the normalized entries shown in Figure~\ref{fig:ECellTotal}. Therefore, from the perspective of background rejection, the energy limit should be set to at least 1 GeV.

Overall, a 4 GeV energy dynamic range is optimal for DarkSHINE ECAL, as it does not impact the accuracy required for energy measurements while still ensuring a great capability of background rejection. The readout electronics in future detectors can be designed based on this standard.

\subsection{Trigger efficiency}

In the high-frequency beam environment of DarkSHINE, the majority of events collected by the detector are background events that do not interest us. These background events can significantly strain data transmission and storage on the backend. However, by implementing an online trigger to directly filter out events that clearly don't match the characteristics of a dark photon signals, we can save a considerable amount of resources. A preliminary design for such a trigger involves summing the energy across selected channels in ECAL. If the summed energy is less than 4 GeV, the event will be saved. A trigger efficiency is defined as the ratio of the number of triggered events, where the summed energy in the trigger region is below 4 GeV, to the total number of events. Table~\ref{tab:Trigger} shows the trigger efficiency across different trigger regions.

\begin{table}[htbp]
\centering
\caption{\label{tab:Trigger}~Table of trigger efficiency across different trigger regions. The trigger efficiency is defined as the ratio of the number of triggered events, where the summed energy in the trigger region is below 4 GeV, to the total number of events. The trigger regions \uppercase\expandafter{\romannumeral 1}, \uppercase\expandafter{\romannumeral 2}, \uppercase\expandafter{\romannumeral 3} and \uppercase\expandafter{\romannumeral 4} correspond to those shown in Figure~\ref{fig:geo}.}
\fontsize{9}{13}\selectfont
    \begin{minipage}[t]{0.5\textwidth}
    \centering
        \begin{tabular}{cccc}
            \toprule
            \makecell[c]{Trigger region} & \makecell[c]{Inclusive} & \makecell[c]{1 MeV dark photon} & \makecell[c]{10 MeV dark photon}\\
            \midrule
            No trigger  & 1M/1M & 100\% & 100\% \\
            \uppercase\expandafter{\romannumeral 1}  & 746/1M & 77.57\% & 91.58\% \\
            \uppercase\expandafter{\romannumeral 1}+\uppercase\expandafter{\romannumeral 2} & 37/1M & 74.05\% & 90.03\% \\
            \uppercase\expandafter{\romannumeral 1}+\uppercase\expandafter{\romannumeral 2}+\uppercase\expandafter{\romannumeral 3} & 6/1M & 73.78\% & 89.9\% \\
            \uppercase\expandafter{\romannumeral 1}+\uppercase\expandafter{\romannumeral 2}+\uppercase\expandafter{\romannumeral 3}+\uppercase\expandafter{\romannumeral 4} & 3/1M & 73.75\% & 89.9\% \\
            \bottomrule
        \end{tabular}
    \end{minipage}%
    \hfill
    \begin{minipage}[t]{0.5\textwidth}
    \centering
        \begin{tabular}{ccc}
            \toprule
            \makecell[c]{Trigger region} & \makecell[c]{100 MeV dark photon} & \makecell[c]{1000 MeV dark photon} \\
            \midrule
            No trigger  & 100\% & 100\%\\
            \uppercase\expandafter{\romannumeral 1}  & 94.86\% & 98.75\%\\
            \uppercase\expandafter{\romannumeral 1}+\uppercase\expandafter{\romannumeral 2}  & 93.67\% & 98.35\%\\
            \uppercase\expandafter{\romannumeral 1}+\uppercase\expandafter{\romannumeral 2}+\uppercase\expandafter{\romannumeral 3} & 93.57\% & 98.32\%\\
            \uppercase\expandafter{\romannumeral 1}+\uppercase\expandafter{\romannumeral 2}+\uppercase\expandafter{\romannumeral 3}+\uppercase\expandafter{\romannumeral 4} & 93.57\% & 98.32\%\\
            \bottomrule
        \end{tabular}
    \end{minipage}
\end{table}

In Table~\ref{tab:Trigger}, with a 4 GeV trigger, most of the inclusive background will be rejected, while dark photon signals, especially those with large masses, will suffer only a small loss. By using only 125 crystals in Region-\uppercase\expandafter{\romannumeral 1}, over 99.9\% of the inclusive backgrounds can be filtered out, which can effectively conserve resources such as bandwidth and storage. When using the sum of the energy from all crystals in the ECAL for triggering, an efficiency of nearly 100\% can be achieved, except for several rare background processes. The trigger strategy shows promise in significantly reducing background events and a small impact on dark photon signals, even when utilizing only the channels in Region-\uppercase\expandafter{\romannumeral 1}. It will serve as an important reference for the electronics design of future detector. 

\section{Energy Resolution}

\subsection{Energy Digitization}
\label{subsec:Digitization}

Accurate simulation of the detector response requires a precise description of digitization effects. Digitization applies a series of realistic effects to the simulation results, mimicking the behavior of an actual detector and bringing the simulation more comparable to real experimental conditions. The energy directly obtained from Geant4, known as the truth energy, represents the ideal energy deposition of a particle in a perfect detection scenario, devoid of any detector effects or measurement errors. Thus, digitization provides a more realistic depiction of detector performance, offering stronger evidence for design.~\cite{Zhu:2024jrf, Li:2023}

The main goal of energy digitization is to parameterize the behavior of each element related to energy measurement. This involves applying smearing based on the preliminary experiments. For the DarkSHINE ECAL, the energy digitization process can be primarily divided into three mian parts: scintillation digitization, SiPM digitization, and ADC digitization. These parts correspond to the scintillator’s light emission and decay, the SiPM response, and the behavior of the readout electronics, respectively, and are based on preliminary experimental measurements.

\subsubsection{Scintillation digitization}

The first step is scintillation digitization. In this part, we sampled the fluctuations that may occur during the generation and attenuation of scintillation light. We set an intrinsic light yield of 30,000 photons per MeV for the LYSO crystal scintillator, assuming a 10\% fluctuation across all scintillators. The scintillation light is proportional to the energy deposited in the crystal and assumed to follow a Poisson distribution. The scintillation light experiences attenuation during propagation within the crystal and detection by the SiPM. The attenuation depends on the photon transportation length and photon detection efficiency (PDE) of the SiPM. The measured light yield helps to determine the amount of light loss during propagation and detection. Although the measured light yield for all crystal-SiPM units shows fluctuations, these can be calibrated to some extent, assumed here to have a 1\% accuracy. The parameters used in this step are shown in Table~\ref{tab:ScinDigi}.

\begin{table}[htbp]
\centering
\caption{\label{tab:ScinDigi} Parameters in scintillation digitization.}
\fontsize{9}{13}\selectfont
    \begin{minipage}[t]{0.5\textwidth}
    \centering
        \begin{tabular}{ccc}
            \toprule
            \makecell[c]{Scintillator} & \makecell[c]{Light yield\\(intrinsic)} & \makecell[c]{Light yield\\ fluctuation} \\
            \midrule
            LYSO & 30000 ph/MeV & 10\% \\
            \bottomrule
        \end{tabular}
    \end{minipage}%
    \hfill
    \begin{minipage}[t]{0.5\textwidth}
    \centering
        \begin{tabular}{cc}
            \toprule
            \makecell[c]{Light yield\\ calibration accuracy} & \makecell[c]{Light yield\\(measured)} \\
            \midrule
            1\% & 150 p.e./MeV \\
            \bottomrule
        \end{tabular}
    \end{minipage}
\end{table}

\subsubsection{SiPM digitization}

For SiPM digitization, a toy Monte Carlo model \cite{zhao2024} simulates the SiPM's response to scintillation light from a LYSO crystal. This model is specifically designed to describe the relationship between the number of photons detected by the SiPM and the number of incident photons. The simulation is based on the HAMAMATSU S14160-3010PS datasheet \cite{S14160-3010PS}, featuring a 3×3 mm$^2$ sensitive area, 10 $\mu$m pixels, and nearly 90,000 pixels in total, with a LYSO crystal measuring 2.5$\times$2.5$\times$4 cm$^3$. The model accounts for the PDE, pixel density, pixel recovery, and crosstalk effects of the SiPM, as well as the scintillation decay time, transmittance, emission spectrum, and absorption spectrum of the LYSO crystal. The parameters used in the SiPM digitization are listed in Table~\ref{tab:SiPMDigi}.

\begin{table}[htbp]
\centering
\caption{\label{tab:SiPMDigi}~Parameters in SiPM digitization.}
\fontsize{9}{13}\selectfont
    \begin{minipage}[t]{0.5\textwidth}
    \centering
        \begin{tabular}{ccc}
            \toprule
            \makecell[c]{Active area \\ (mm$^2$)} & \makecell[c]{Pixel pitch \\ ($\mu$m)} & \makecell[c]{Pixel number} \\
            \midrule
            3.0$\times$3.0 & 10 & 89984 \\
            \bottomrule
        \end{tabular}
    \end{minipage}%
    \hfill
    \begin{minipage}[t]{0.5\textwidth}
    \centering
        \begin{tabular}{ccc}
            \toprule
            \makecell[c]{PDE \\ ($\lambda=\lambda_p$)} & \makecell[c]{Fill factor} & \makecell[c]{Gain fluctuation}\\
            \midrule
            18\% & 31\% & 10\%\\
            \bottomrule
        \end{tabular}
    \end{minipage}
\end{table}

\begin{figure}[htbp]
\centering
\includegraphics[width=0.8\linewidth]{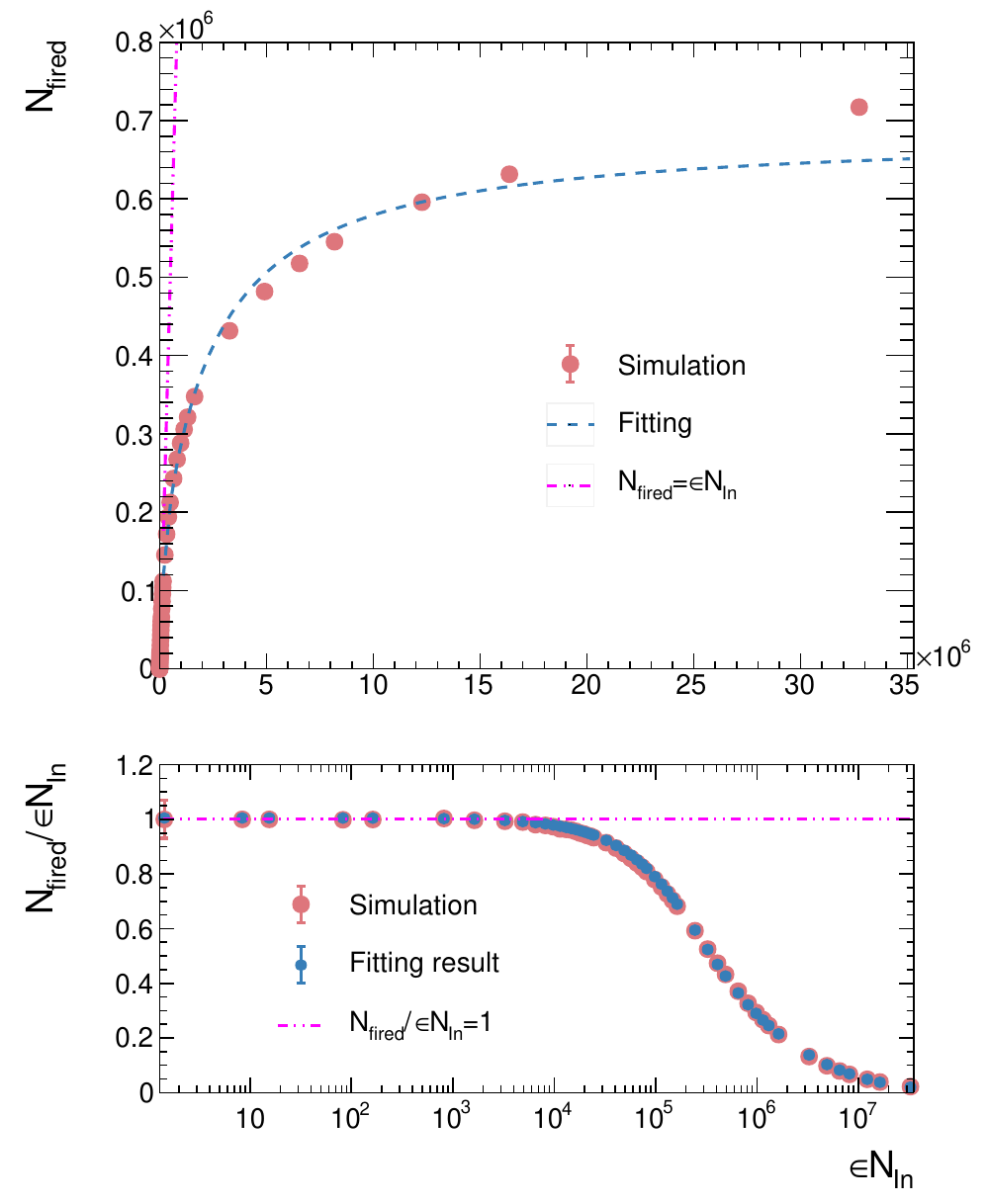}
    \caption{\label{fig:SiPMResponse}~The simulated response of the SiPM to scintillation light from a 2.5$\times$2.5$\times$4 cm$^3$ LYSO crystal is shown. This SiPM-LYSO setup is identical to that used in the DarkSHINE ECAL. Performance parameters of the SiPM, including PDE, fill factor, pixel pitch, pixel counts, and recovery time, are referenced from the HAMAMATSU S14160-3010PS datasheet \cite{S14160-3010PS}. The detected photons exhibit a time structure influenced by the scintillation decay time of LYSO and the geometric effects of the crystal, which is considered in the simulation. The horizontal axis represents the effective photon counts, calculated as the product of the PDE ($\epsilon$) and the number of incident photons ($N_{In}$). The upper figure displays the variation in the number of photons detected by the SiPM ($N_{fired}$) as a function of the effective photon counts. The lower figure illustrates the degree to which the SiPM’s response deviates from linearity. It can be observed that for effective photon counts below 10,000, the SiPM’s response remains largely linear. However, as the effective photon counts exceeds 10,000, the response begins to deviate from linearity and gradually saturates.
    }
\end{figure}

As shown in Figure~\ref{fig:SiPMResponse}, $N_{fired}$ represents the number of photons detected by the SiPM, $\epsilon$ denotes the PDE of SiPM, and $N_{In}$ refers to the number of photons incident on the SiPM. The SiPM exhibits a nearly linear response when the effective photon count, calculated as the product of the PDE ($\epsilon$) and the number of incident photons ($N_{In}$), is below 10,000. As the effective photon counts increases, the response gradually deviates from linearity and approaches saturation. A formula from \cite{kotera2016describing} is employed to fit the SiPM's response, which aids in modeling its behavior for varying numbers of incident photons, and provides a means to correct for saturation effects. 

In the digitization of SiPM, the number of photoelectrons in the scintillation digitization is used as the input for the fitting function shown in Figure~\ref{fig:SiPMResponse}, resulting in a response that includes the non-linear effects of the SiPM. The fluctuations during this process can be also obtained, which is approximately proportional to the square root of the effective photon counts. Furthermore, the inverse of this fitting function can be applied to correct the non-linear response of SiPM, while this correction will inevitably introduce fluctuations once again. \cite{PIEMONTE20192, NAGY201444}

\subsubsection{ADC digitization}

The behavior of the readout electronics is simulated in the ADC digitization process. Table~\ref{tab:ADCDigi} shows the parameters used in ADC digitization. Here, we assume that the charge produced by the SiPM is recorded by a 12-bit multi-channel analyzer with a total of 4096 ADC counts. The gain is defined as the number of ADC counts output per single photoelectron signal. There are three gain modes to expand the dynamic range of energy measurement: high gain, medium gain, and low gain. The energy dynamic range for high gain is approximately 0 to 2.7 MeV, primarily designed for energy calibration using radioactive sources. The maximum energy measured by the designed electronics is approximately 5461 MeV, which is higher than the 4 GeV result shown in Section~\ref{sec:EnergyDis}. This increase is due to a series of smearing effects during digitization, which can result in energies greater than the original values. In high gain and medium gain modes, when the ADC exceeds 4000, it will automatically switch to the next gain level. A 3-ADC DAQ noise was set for each gain mode, representing the intrinsic noise of the data acquisition system. The SiPM noise is set to 1 ADC at high gain mode, and varies with gain. The digitization model characterizes the SiPM noise as the average dark noise present in each event's waveform. This accounts for the scintillation decay time of LYSO and the dark count rate of the SiPM. Additionally, a 1\% calibration accuracy was also assumed during photoelectron measurement. 

\begin{table}[htbp]
\centering
\caption{\label{tab:ADCDigi} Parameters in ADC digitization.}
\fontsize{9}{13}\selectfont
    \begin{minipage}[t]{0.5\textwidth}
    \centering
        \begin{tabular}{cccc}
            \toprule
            \makecell[c]{Modes} & \makecell[c]{Gain} & \makecell[c]{ADC/p.e.} & \makecell[c]{Charge calibration \\ accuracy}\\
            \midrule
            High gain   & $\times$2000   & 10 & 1\%\\
            Medium gain & $\times$40  & 0.2 & 1\%\\
            Low gain   & $\times$1 & 0.005 & 1\%\\
            \bottomrule
        \end{tabular}
    \end{minipage}%
    \hfill
    \begin{minipage}[t]{0.5\textwidth}
    \centering
        \begin{tabular}{cccc}
            \toprule
            \makecell[c]{SiPM noise} & \makecell[c]{DAQ noise} & \makecell[c]{Switching point} & \makecell[c]{Energy range} \\
            \midrule
            1 ADC & 3 ADC & 4000 ADC & 0-2.7 MeV \\
            0.02 ADC & 3 ADC & 4000 ADC & 2.7-133.3 MeV \\
            0.0005 ADC & 3 ADC & -        & 133.3-5461 MeV \\
            \bottomrule
        \end{tabular}
    \end{minipage}
\end{table}

\subsection{Single channel performance}

\begin{figure}[htbp]
\centering  %图片全局居中
\subfigure[]{
\includegraphics[width=0.45\textwidth]{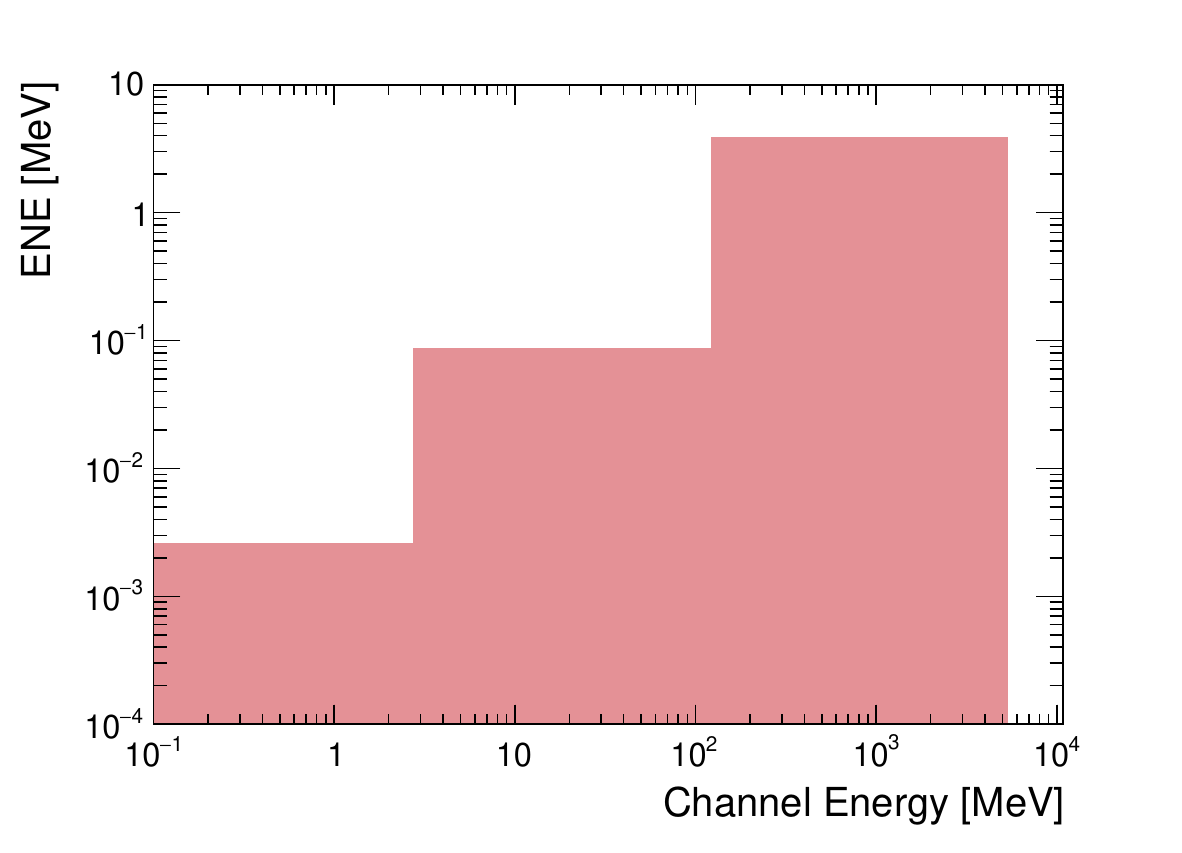}}
\subfigure[]{
\includegraphics[width=0.45\textwidth]{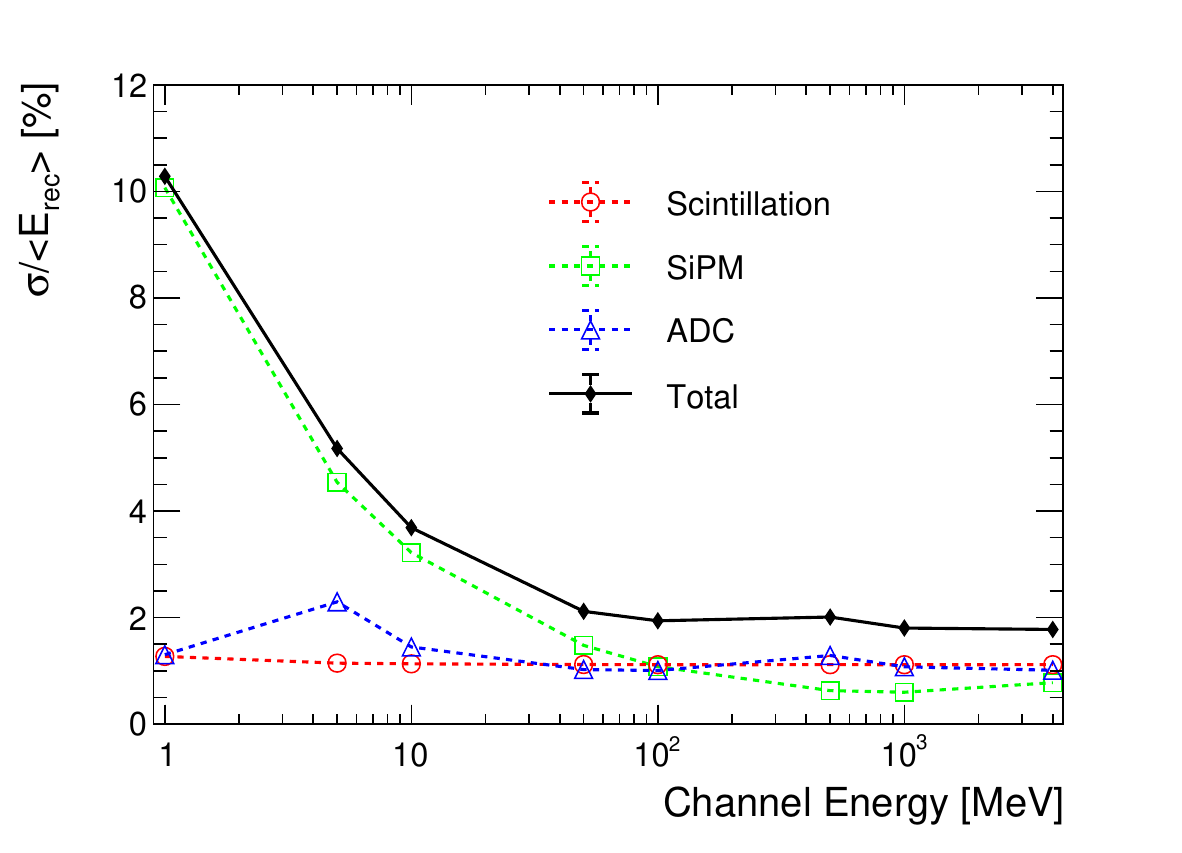}}
\caption{\label{fig:DigiSize}~(a) Equivalent noise energy (ENE) of a single channel. The x-axis represents the input energy in digitization model, while the y-axis represents ENE. The three bins in the figure correspond to the three gain modes, each with a fixed ENE value that is independent of input energy. The width of each bin represents the dynamic range of the energy measurement at that gain.
(b) Energy resolution of a single channel. The x-axis represents the input energy in digitization model, and y-axis indicates the channel's energy resolution, calculated as the ratio of the sigma to the mean value from a Gaussian fit of the output energy distribution. At energies below 100 MeV, the primary contributor to resolution is the SiPM digitization, since it depends on efficiencies of light collection and photon detection, which introduced larger fluctuations in monte carlo sampling than intrinsic light output and ADC noise. As energy increases, the scintillation digitization and ADC digitization become more dominant, mainly due to the constant terms reflecting calibration accuracy.}
\end{figure}

After digitization, the ideal energy deposited in the crystal within Geant4 is transformed into a signal that more closely represents the actual response of the detector. The Equivalent Noise Energy (ENE) for the electronics of a single channel digitization is shown in Figure~\ref{fig:DigiSize}(a). The ENE is calculated using the following formula:
\begin{equation}
ENE = \frac{Noise}{ADC_{p.e.} \times LY_{mea}}
\end{equation}
where $Noise$ is the number of ADC counts of noise, representing the sum of SiPM noise and DAQ noise, as listed in Table~\ref{tab:ADCDigi}. $ADC_{p.e.}$ is the number of ADC counts per photoelectron, also provided in Table~\ref{tab:ADCDigi}. $LY_{mea}$ is the number of photoelectrons detected by the SiPM coupled to the crystal when the crystal absorbs 1 MeV of energy. In our setup, $LY_{mea}$ is set to 150 p.e./MeV (Table~\ref{tab:ScinDigi}). 

In Figure~\ref{fig:DigiSize}(a), three distinct ENE regions correspond to different gain modes. Since ENE represents the electronics noise and is independent of the input signal, it remains constant across the energy range covered by each electronics gain. However, at the transition points between these ranges, the ENE exhibits jumps, as the gain changes as well as $Noise$ and $ADC_{p.e.}$. 

After incorporating the three digitization effects, the single-channel energy resolution is shown in Figure~\ref{fig:DigiSize}(b). We individually simulated the digitization process for each component to obtain its resolution and also simulated the process with all three components present to determine the overall resolution. The scintillation digitization consists of a Poisson sampling based on the intrinsic light yield of LYSO crystal, along with a series of Gaussian samplings. As a result, this part of the energy resolution decreases with increasing incident energy. But, due to the high intrinsic light yield of the crystal, this change is not very noticeable in Figure~\ref{fig:DigiSize}(b) and is instead dominated by the constant term introduced by the Gaussian sampling, which reflects the calibration accuracy. The resolution of SiPM digitization also decreases with increasing energy. However, since its statistical sampling is based on the number of detected photoelectrons, which is significantly lower than the number of generated scintillation photons due to attenuation and the PDE of SiPM, the resolution of SiPM digitization is worse than that of scintillation digitization. Moreover, applying a non-linearity correction further degrades the resolution, as the correction process introduces additional fluctuations. For ADC digitization, the energy resolution shows two jumps at 5 MeV and 500 MeV, caused by an increase in ENE due to gain switching. Additionally, a constant term appears in the high-energy region, reflecting the calibration accuracy assumed in the model. Overall, at lower energies (less than 100 MeV), the resolution of the SiPM digitization is the main contributor. At higher energies, contributions from the scintillation digitization and ADC digitization become more significant. 

The energy resolutions in Figure~\ref{fig:DigiSize}(b) were obtained based on parameters in Table~\ref{tab:ScinDigi}, \ref{tab:SiPMDigi} and \ref{tab:ADCDigi}, which referred to the preliminary experimental measurements. Optimizing the LYSO-SiPM units, such as improving light collection efficiency to get a larger light output, can enhance the channel resolution.

\subsection{ECAL performance}

\begin{figure}[htbp]
\centering
\includegraphics[width=0.8\linewidth]{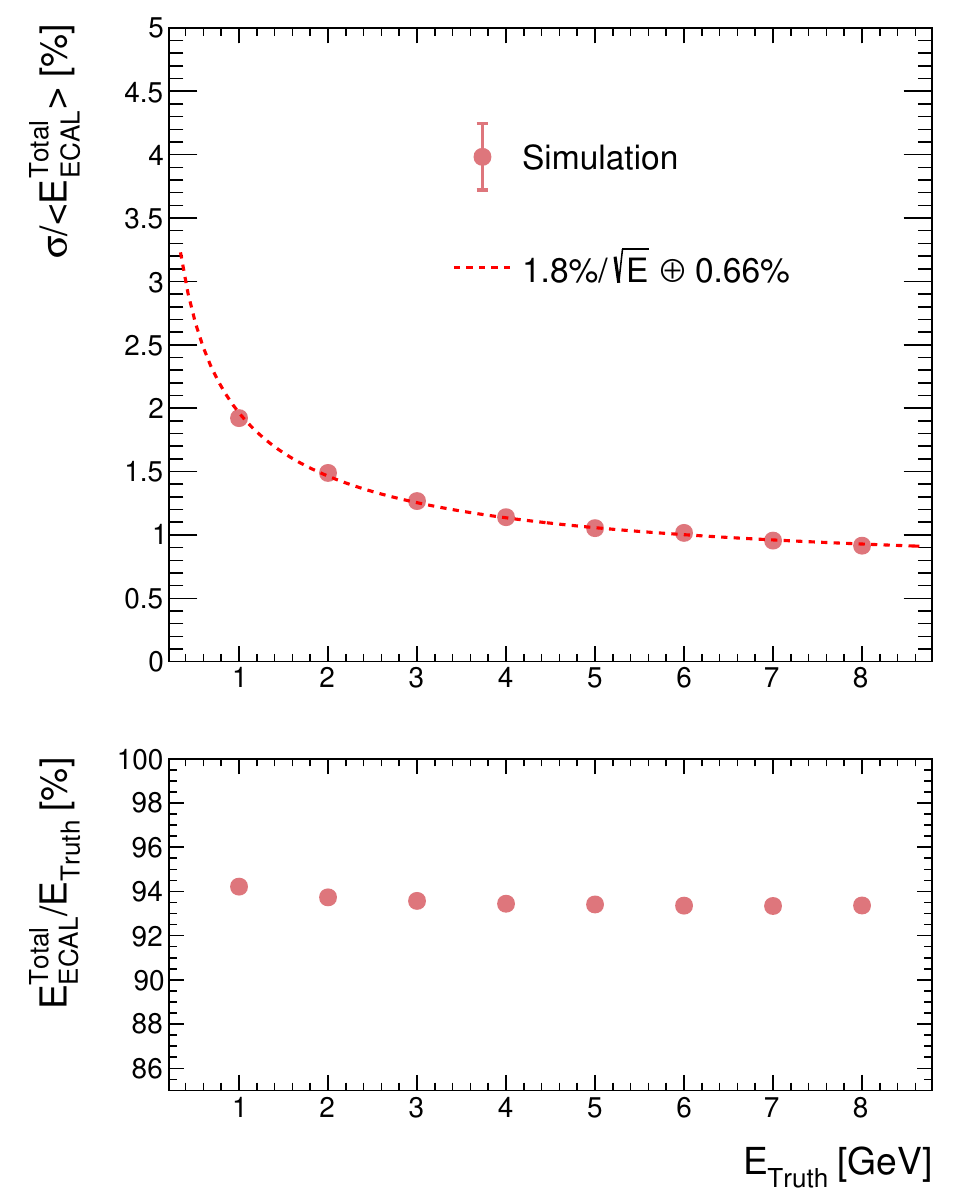}
    \caption{\label{fig:ECALPerformance}~Energy resolution and energy containment of the ECAL for 1–8 GeV incident electrons, without the target and other detectors. The x-axis represents the energy of the incident electrons. The top figure displays the ECAL's energy resolution, with the fitted stochastic term achieving better than 2\%. The bottom figure shows the energy containment, defined as the ratio of the energy deposited in the ECAL to the incident energy.
    }
\end{figure}

The performance of the full detector can be determined by applying digitization to each channel. Figure~\ref{fig:ECALPerformance} show the energy resolution and energy containment of the ECAL. $E_{Truth}$ is the energy of incident electrons. In simulation, the ECAL is configured to consist of 21$\times$21$\times$11 LYSO crystals, each with a volume of 2.5$\times$2.5$\times$4 cm$^3$, consistent with the results from Section~\ref{sec:VolOpt}. The target and other detectors are not included. The incident particles are electrons with energies ranging from 1 to 8 GeV, hitting the center of ECAL's front face within a circular area with a radius of 3 cm. 

The energy resolution is plotted at the top of Figure~\ref{fig:ECALPerformance}. After digitization, the statistical term of ECAL's energy resolution remains better than 2\%, thanks to the very high intrinsic light yield of LYSO crystal. The high energy resolution will allow for more precise measurement of recoil electron energy and enhance the reliability in distinguishing between signal and background. The overall energy resolution of the ECAL is typically better than that of a single channel (Figure~\ref{fig:DigiSize}(b)) due to statistical averaging, where uncertainties from individual channels tend to cancel out, and energy sharing across multiple channels, which reduces the impact of measurement errors. Additionally, variations in the response of different channels are compensated when combined, improving the overall resolution.

The bottom of Figure~\ref{fig:ECALPerformance} shows the energy containment of ECAL, which is defined as the ratio of deposited energy in ECAL to the incident energy. The energy containment is less than 100\%, mainly because a small portion of the energy is deposited in passive materials outside the crystals, such as the reflective films, PCB, and support structures, which can not be detected.

Since the performance of the ECAL is influenced by the digitization model, which is based on preliminary experimental results from a specific setup, optimizing the detector configuration could change the performance. However, the key objective is to develop a tool that can more accurately evaluate the detector's performance, making it closer to that of a real detector.

\section{Radiation damage}

Given the high-energy and high-frequency beam environment, the ECAL, particularly its central region, is subjected to significant radiation dose that may degrade its performance. Therefore, the radiation damage to the ECAL must be evaluated, focusing on crystal damage primarily from ionizing energy loss and silicon sensor damage from non-ionizing energy loss. Simulations were conducted using Geant4 to estimate the radiation damage to crystals and silicon sensors under $3\times10^{14}$ electrons-on-target events, corresponding to one year of operation at a 10 MHz repetition rate. 

\begin{figure}[htbp]
\centering
\subfigure[]{
\includegraphics[width=0.45\textwidth]{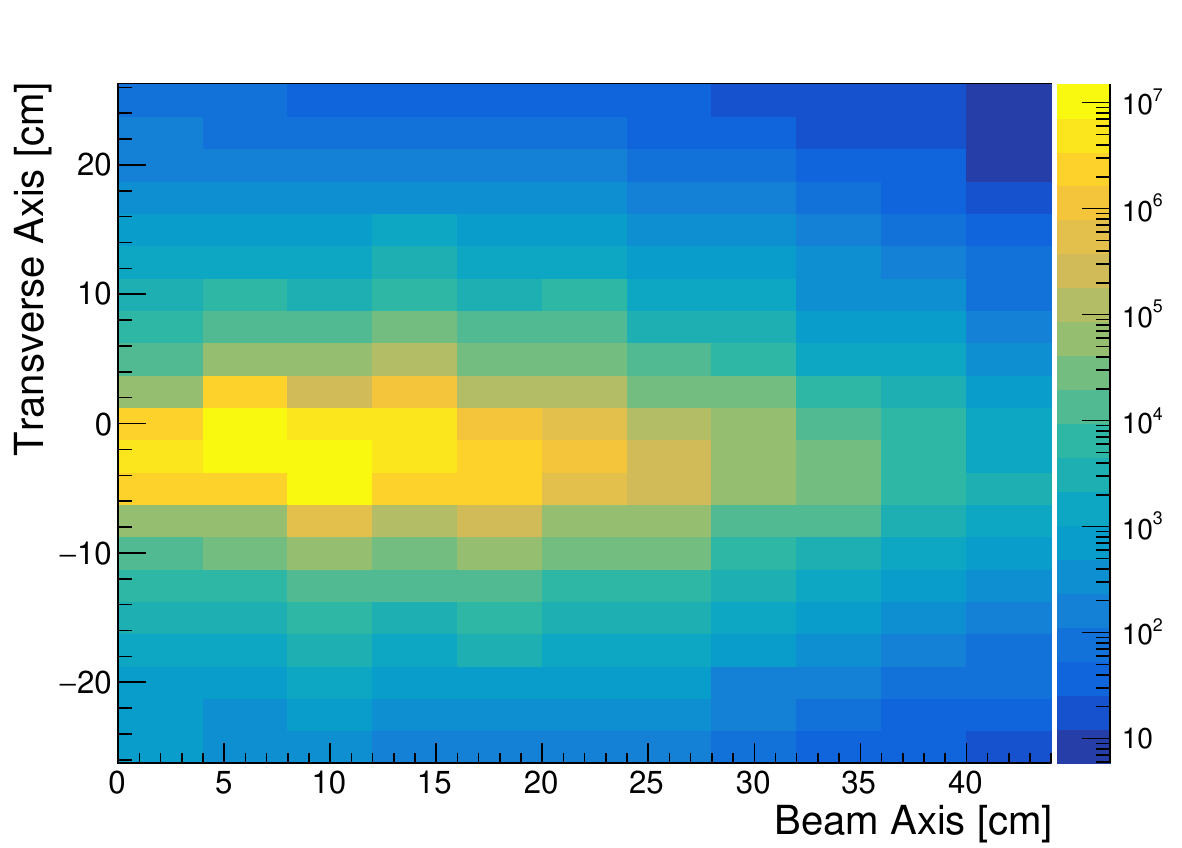}}
\subfigure[]{
\includegraphics[width=0.45\textwidth]{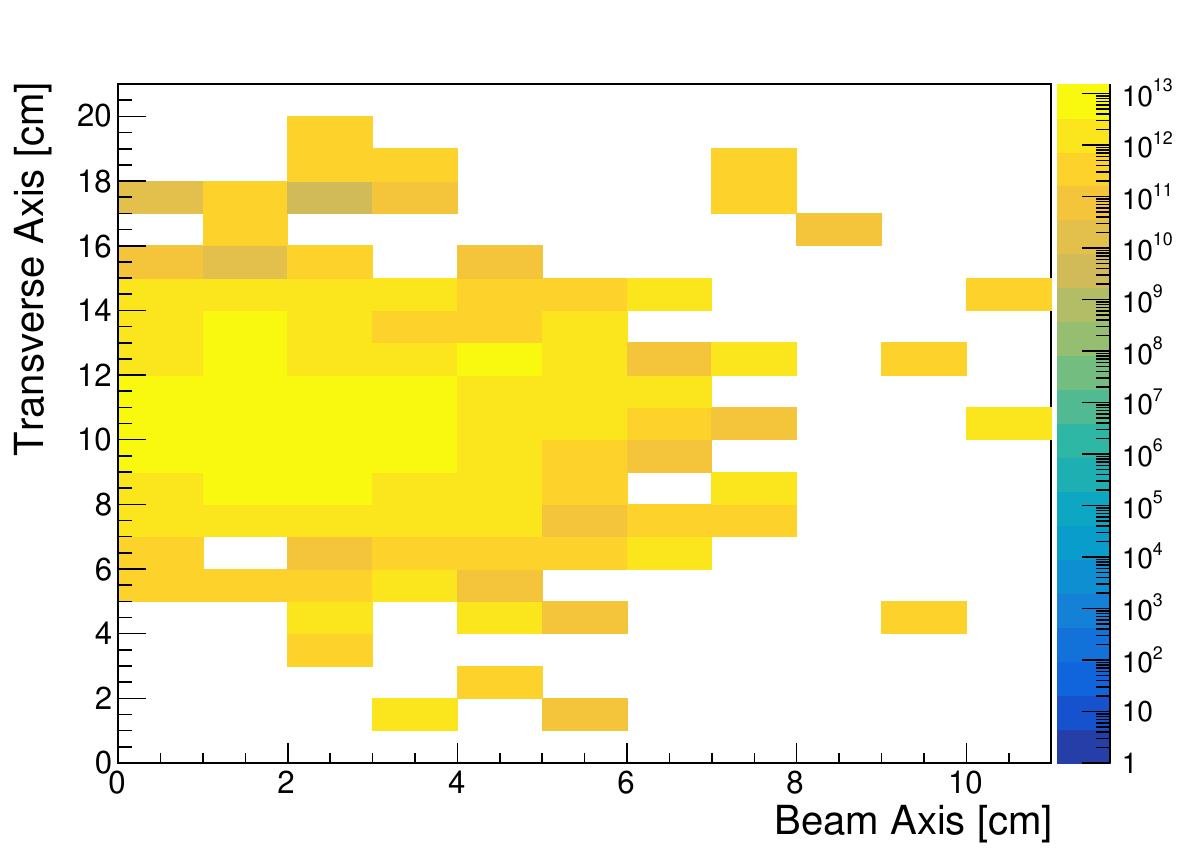}}
\caption{\label{fig:RadiationDamage}~Distribution of radiation damage in the ECAL region under $3\times10^{14}$ electrons-on-target events, corresponding to one year of operation at a 10 MHz repetition rate. The two distributions illustrate the radiation damage along the ECAL symmetry plane in beam direction. In figures, the horizontal axis represents the beam direction, while the vertical axis corresponds to the transverse directions. Each segment indicates the position of a crystal. (a) Total Ionizing Dose (TID) absorbed by crystals, with a maximum value of $10^7$ rad. (b) Non-Ionizing Energy Loss (NIEL) in silicon sensors, expressed as the equivalent 1 MeV neutron flux, with a maximum value of $10^{13}$ per square centimeter.}
\end{figure}

Radiation damage to crystals mainly results from ionizing energy loss of incident particles, which can be evaluated by the Total Ionizing Dose (TID). The TID in crystal is defined as the ionizing energy absorbed per unit mass of crystal. The average ionizing energy loss absorbed by each crystal in the ECAL per event was calculated based on one million electrons-on-target events. This average value was then multiplied by $3\times10^{14}$ to estimate the TID for each crystal after one year of operation. For the crystal absorb the maximum dose, the value of TID is about $10^7$ rad (Figure~\ref{fig:RadiationDamage}(a)). Most inorganic scintillators commonly used in high-energy physics detectors, like CsI, BGO, and PWO, lose significant light yield after such a dose. However, LYSO exhibits only a small reduction in light yield\cite{Zhu:2019ihr}, meeting the radiation resistance requirements for the DarkSHINE ECAL.

Radiation damage to silicon sensors mainly results from Non-Ionizing Energy Loss (NIEL), typically expressed using the equivalent 1 MeV neutron flux. To obtain the flux on each sensor, the average NIEL per event for each sensor was first calculated through simulation, referred to as $E_1$. Then, the average NIEL for a 1 MeV neutron passing through a single sensor was simulated, referred to as $E_2$. Finally, the ratio $E_1/E_2$ provides the equivalent 1 MeV neutron flux for each sensor. In the DarkSHINE ECAL, the equivalent 1 MeV neutron flux on silicon sensors in the most heavily irradiated area is about $10^{13}$ per square centimeter (Figure~\ref{fig:RadiationDamage}(b)). This significant radiation could cause the dark current of general sensors to increase by several orders of magnitude, rendering them unusable. Therefore, silicon sensors with excellent radiation resistance are required for our experiment.\cite{ULYANOV2020164203, SANCHEZMAJOS2009506, PREGHENELLA2023167661}

\section{Conclusion}

In this study, we presented the design and optimization of a LYSO crystal ECAL for the DarkSHINE experiment, aimed at detecting dark photons, potential candidates for dark force mediators. Through comprehensive simulations, we optimized the ECAL’s structure and evaluated its key performance metrics.

The final ECAL design consists of 21$\times$21$\times$11 LYSO crystals, each measuring 2.5$\times$2.5$\times$4 cm$^3$, arranged in a staggered configuration to enhance detection efficiency. Both the transverse and longitudinal dimensions of ECAL were optimized to achieve high signal collection while maintaining cost efficiency. A 4 GeV energy dynamic range was established for each channel to ensure accurate energy measurements without saturation, which is crucial for background rejection and precise signal detection.

A dedicated digitization model parameterizing the scintillation, SiPM, and ADC behaviors was developed to provide a realistic representation of the detector’s performance. This model allowed us to study the energy resolution and containment of the ECAL, demonstrating that the statistical term of the energy resolution can be better than 2\%, thanks to the high intrinsic light yield of the LYSO crystals.

Radiation damage was also assessed. The TID absorbed by the crystals and the non-ionizing energy loss in the silicon sensors were estimated for one year of operation at a 10 MHz repetition rate. Due to the small light yield reduction under high radiation doses, LYSO crystal is an ideal scintillator material for the DarkSHINE experiment. The selected silicon sensors for the ECAL should also have excellent resistance to radiation-induced damage, which is critical for maintaining detector performance in the high-radiation environment anticipated for the experiment.


\section{Bibliography}

\begin{thebibliography}{99}

\bibitem{Hooper:2007kb}
D.~Hooper, D.~P.~Finkbeiner and G.~Dobler,
Possible evidence for dark matter annihilations from the excess microwave emission around the center of the Galaxy seen by the Wilkinson Microwave Anisotropy Probe.
Phys. Rev. D \textbf{76}, 083012 (2007).
doi:10.1103/PhysRevD.76.083012

\bibitem{Clowe:2006}
D.~Clowe, M.~Bradac, A.~H.~Gonzalez, et al.,
A direct empirical proof of the existence of dark matter.
Astrophys. J. Lett. \textbf{648}, L109-L113 (2006).
doi:10.1086/508162,

\bibitem{Du:2021jcj}
Y.~Du, F.~Huang, H.~L.~Li et al.,
Revisiting dark matter freeze-in and freeze-out through phase-space distribution. 
JCAP \textbf{04}, 012 (2022). 
doi:10.1088/1475-7516/2022/04/012

\bibitem{XENON:2023cxc}
E.~Aprile, K.~Abe, F.~Agostini, et al.,
First Dark Matter Search with Nuclear Recoils from the XENONnT Experiment. 
Phys. Rev. Lett. \textbf{131}, 041003 (2023). 
doi:10.1103/PhysRevLett.131.041003

\bibitem{PandaX-4T:2021bab}
Y.~Meng, Z.~Wang, Y.~Tao, et al.,
Dark Matter Search Results from the PandaX-4T Commissioning Run. 
Phys. Rev. Lett. \textbf{127}, 261802 (2023). 
doi:10.1103/PhysRevLett.127.261802

\bibitem{Wang_2021}
Y.~Wang, Z.~Zeng, Q.~Yue, et al.,
First experimental constraints on WIMP couplings in the effective field theory framework from CDEX.
Sci. China. Phys. Mech. Astron. \textbf{64}, 281011 (2021). 
doi:10.1007/s11433-020-1666-8

\bibitem{PhysRevLett.131.041002}
J.~Aalbers, D.~S.~Akerib, C.~W.~Akerlof, et al.,
First Dark Matter Search Results from the LUX-ZEPLIN (LZ) Experiment.
Phys. Rev. Lett. \textbf{131}, 041002 (2023). 
doi:10.1103/PhysRevLett.131.041002

\bibitem{Giovacchini:2020vxz}
F.~Giovacchini, J.~Casaus, A.~Oliva, et al.,
The AMS-02 RICH detector: Status and physics results.
Nucl. Instrum. Meth. A \textbf{952}, 161797 (2020). 
doi:10.1016/j.nima.2019.01.024

\bibitem{Kyratzis:2022gvg}
D.~Kyratzis,
Results overview from the DAMPE space mission in orbit.
PoS \textbf{PANIC2021}, 310 (2022).
doi:10.22323/1.380.0310

\bibitem{Giagu:2019fmp}
G.~Stefano,
WIMP Dark Matter Searches With the ATLAS Detector at the LHC.
Front. in Phys. \textbf{7}, 75 (2019).
doi:10.3389/fphy.2019.00075

\bibitem{Griest:1989wd}
K.~Griest and M.~Kamionkowski,
Unitarity Limits on the Mass and Radius of Dark Matter Particles.
Phys. Rev. Lett. \textbf{64}, 615 (1990).
doi:10.1103/PhysRevLett.64.615

\bibitem{Ho:2012ug}
C.~M.~Ho and R.~J.~Scherrer,
Limits on MeV Dark Matter from the Effective Number of Neutrinos.
Phys. Rev. D \textbf{87}, 023505 (2013).
doi:10.1103/PhysRevD.87.023505

\bibitem{Steigman:2013yua}
G.~Steigman,
Equivalent Neutrinos, Light WIMPs, and the Chimera of Dark Radiation.
Phys. Rev. D \textbf{87}, 103517 (2013).
doi:10.1103/PhysRevD.87.103517

\bibitem{Boehm:2013jpa}
C.~Boehm, M.~J.~Dolan and C.~McCabe,
A Lower Bound on the Mass of Cold Thermal Dark Matter from Planck.
JCAP \textbf{08}, 041 (2013).
doi:10.1088/1475-7516/2013/08/041

\bibitem{Nollett:2013pwa}
K.~M.~Nollett and G.~Steigman,
BBN And The CMB Constrain Light, Electromagnetically Coupled WIMPs.
Phys. Rev. D \textbf{89}, 083508 (2014).
doi:10.1103/PhysRevD.89.083508

\bibitem{Nollett:2014lwa}
K.~M.~Nollett and G.~Steigman,
BBN And The CMB Constrain Neutrino Coupled Light WIMPs.
Phys. Rev. D \textbf{91}, 083505 (2015).
doi:10.1103/PhysRevD.91.083505

\bibitem{Serpico:2004nm}
P.~D.~Serpico and G.~G.~Raffelt,
MeV-mass dark matter and primordial nucleosynthesis.
Phys. Rev. D \textbf{70}, 043526 (2004).
doi:10.1103/PhysRevD.70.043526

\bibitem{Billard:2021uyg}
J.~Billard, M.~Boulay, S.~Cebri\'an, et al.,
Direct detection of dark matter\textemdash{}APPEC committee report*.
Rept. Prog. Phys. \textbf{85}, 056201 (2022).
doi:10.1088/1361-6633/ac5754

\bibitem{Holdom:1985ag}
B.~Holdom,
Two U(1)'s and Epsilon Charge Shifts.
Phys. Lett. B \textbf{166}, 196--198 (1986).
doi:10.1016/0370-2693(86)91377-8

\bibitem{Foot:1991kb}
R.~Foot and X.~G.~He,
Comment on Z Z-prime mixing in extended gauge theories.
Phys. Lett. B \textbf{267}, 509--512 (1991).
doi:10.1016/0370-2693(91)90901-2

\bibitem{Fuyuto:2019vfe}
K.~Fuyuto, X.~G.~He, G.~Li, et al.,
CP-violating Dark Photon Interaction.
Phys. Rev. D \textbf{101}, 075016 (2020).
doi:10.1103/PhysRevD.101.075016

\bibitem{Choi:2020pyy}
G.~Choi, T.~T.~Yanagida and N.~Yokozaki,
A model of interacting dark matter and dark radiation for H$_{0}$ and $\sigma_{8}$ tensions.
JHEP \textbf{01}, 127 (2021).
doi:10.1007/JHEP01(2021)127

\bibitem{Cheng_2022}
Y.~Cheng, X.~G.~He, M.~Ramsey-Musolf, et al.,
CP violating dark photon kinetic mixing and type-III seesaw model.
Phys. Rev. D \textbf{105}, 095010 (2022).
doi:10.1103/physrevd.105.095010

\bibitem{Banerjee:2019pds}
D.~Banerjee, V.~E.~Burtsev, A.~G.~Chumakov, et al.,
Dark matter search in missing energy events with NA64.
Phys. Rev. Lett. \textbf{123}, 121801 (2019).
doi:10.1103/PhysRevLett.123.121801

\bibitem{2024138378}
H.~Abreu, J.~Anders, C.~Antel, et al. [FASER Collaboration],
Search for dark photons with the FASER detector at the LHC.
Phys. Lett. B \textbf{848}, 138378 (2024).
doi:10.1016/j.physletb.2023.138378

\bibitem{PhysRevLett.130.071804}
F.~Abudin\'en, I.~Adachi, L.~Aggarwal, et al.,
Search for a Dark Photon and an Invisible Dark Higgs Boson in ${\ensuremath{\mu}}^{+}{\ensuremath{\mu}}^{\ensuremath{-}}$ and Missing Energy Final States with the Belle II Experiment.
Phys. Rev. Lett. \textbf{130}, 071804 (2023).
doi:10.1103/PhysRevLett.130.071804

\bibitem{PhysRevD.100.115016}
Y.~Zhang, W.~T.~Zhang, M.~Song, et al.,
Probing invisible decay of a dark photon at BESIII and a future Super Tau Charm Factory via monophoton searches.
Phys. Rev. D \textbf{100}, 115016 (2019).
doi:10.1103/PhysRevD.100.115016

\bibitem{Prasad:2019ris}
V.~Prasad,
Dark matter/ new physics searches at BESIII.
PoS \textbf{ALPS2019}, 030 (2020).
doi:10.22323/1.360.0030

\bibitem{Fabbrichesi_2021}
M.~Fabbrichesi, E.~Gabrielli and G.~Lanfranchi,
The Physics of the Dark Photon: A Primer.
Springer, (2021).
doi:10.1007/978-3-030-62519-1

\bibitem{LDMX:2018cma}
T.~\r{A}kesson et al. [LDMX],
Light Dark Matter eXperiment (LDMX).
[arXiv:1808.05219 [hep-ex]].

\bibitem{LDMX:2019gvz}
T.~\r{A}kesson et al. [LDMX],
A High Efficiency Photon Veto for the Light Dark Matter eXperiment.
JHEP \textbf{04}, 003 (2020).
doi:10.1007/JHEP04(2020)003

\bibitem{Berlin:2018bsc}
A.~Berlin, N.~Blinov, G.~Krnjaic et al.,
Dark Matter, Millicharges, Axion and Scalar Particles, Gauge Bosons, and Other New Physics with LDMX.
Phys. Rev. D \textbf{99}, 075001 (2019).
doi:10.1103/PhysRevD.99.075001,

\bibitem{Chen:2022liu}
J.~Chen, J.~Y.~Chen, J.~F.~Chen, et al.,
Prospective study of light dark matter search with a newly proposed DarkSHINE experiment. 
Sci. China Phys. Mech. Astron. \textbf{66}, 211062 (2023). 
doi:10.1007/s11433-022-1983-8

\bibitem{SLi}
S.~Li, 
Dark SHINE — a Dark Photon search initiative at SHINE facility.  
Proceeding for 31st International Symposium on Lepton Photon Interactions at High Energies. doi:10.5281/zenodo.8373963

\bibitem{Wan:2022het}
J.~Wan, Y.~Leng, B.~Gao, et al.,
Simulation of wire scanner for high repetition free electron laser facilities. 
Nucl. Instrum. Meth. A \textbf{1026}, 166200 (2022). 
doi:10.1016/j.nima.2021.166200

\bibitem{Zhao:2017ood}
Z.~T.~Zhao, C.~Feng and K.~Q.~Zhang,
Two-stage EEHG for coherent hard X-ray generation based on a superconducting linac. 
Nucl. Sci. Tech. \textbf{28}, 117 (2017). 
doi:10.1007/s41365-017-0258-z

\bibitem{1239590}
T.~Kimble, M.~Chou and B.~H.~T.~Chai,
Scintillation properties of LYSO crystals. 
IEEE \textbf{3}, 1434-1437 (2002). 
doi:10.1109/NSSMIC.2002.1239590

\bibitem{Butler:2019rpu}
J.~N.~Butler, [CMS Collaboration],
A MIP Timing Detector for the CMS Phase-2 Upgrade. 
CERN, CERN-LHCC-2019-003, CMS-TDR-020 (2019)

\bibitem{Kalinnikov:2023coj}
V.~Kalinnikov, E.~Velicheva and A.~Rozhdestvensky,
Measurement of the LYSO:Ce and LYSO:Ce,Ca Scintillator Response for the Electromagnetic Calorimeter of the COMET Experiment. 
Phys. Part. Nucl. Lett. \textbf{20}, 995--1001 (2023). 
doi:10.1134/S1547477123050412

\bibitem{KLANNER201936}
R.~Klanner,
Characterisation of SiPMs. 
Nucl. Instrum. Meth. A \textbf{926}, 36--56 (2019). 
doi:10.1016/j.nima.2018.11.083

\bibitem{SIMON201985}
F.~Simon,
Silicon photomultipliers in particle and nuclear physics. 
Nucl. Instrum. Meth. A \textbf{926}, 85--100 (2019). 
doi:10.1016/j.nima.2018.11.042

\bibitem{Yu:2024quo}
X.~Z.~Yu, X.~Y.~Wang, W.~H.~Ma, et al.,
Production and test of sPHENIX W/SciFiber electromagnetic calorimeter blocks in China,
Nucl. Sci. Tech. \textbf{35}, no.8, 145 (2024)
doi:10.1007/s41365-024-01517-y

\bibitem{Xu:2024btf}
Y.~Xu, Y.~S.~Ning, Z.~Z.~Qin, et al.,
Development of a scintillating-fiber-based beam monitor for the coherent muon-to-electron transition experiment,
Nucl. Sci. Tech. \textbf{35}, no.4, 79 (2024)
doi:10.1007/s41365-024-01442-0

\bibitem{GEANT4:2002zbu}
S.~Agostinelli, J.~Allison, K.~Amako, et al.,
GEANT4--a simulation toolkit. 
Nucl. Instrum. Meth. A \textbf{506}, 250--303 (2003). 
doi:10.1016/S0168-9002(03)01368-8

\bibitem{Zhu:2023}
J.~Zhu, X.~Peng, S.~Luo, et al.,
Performance of the electromagnetic calorimeter module in the NICA-MPD based on Geant4,
Nucl. Sci. Tech. \textbf{46}, no.12, 120202 (2023)
doi:10.11889/j.0253-3219.2023.hjs.46.120202

\bibitem{Izaguirre:2014bca}
E.~Izaguirre, G.~Krnjaic, P.~Schuster, et al.,
Testing GeV-Scale Dark Matter with Fixed-Target Missing Momentum Experiments. 
Phys. Rev. D \textbf{91}, 094026 (2015). 
doi:10.1103/PhysRevD.91.094026

\bibitem{Wang2024}
Z.~Wang, R.~Yuan, H.~Q.~Liu, et al.,
A Design of Hadronic Calorimeter for DarkSHINE Experiment. 
Nucl. Sci. Tech. \textbf{35}, 148 (2024). 
doi:10.1007/s41365-024-01502-5

\bibitem{guo2024}
Y.~H.~Guo, S.~Li, K.~Liu, et al.,
Design of High-speed readout electronics for the DarkSHINE electromagnetic calorimeter. 
[arXiv:2407.20723 [physics.ins-det]]

\bibitem{Zhu:2024jrf}
J.~Y.~Zhu, Y.~Z.~Su, H.~B.~Yang, et al.,
Design and prototyping of the readout electronics for the transition radiation detector in the high energy cosmic radiation detection facility,
Nucl. Sci. Tech. \textbf{35}, no.4, 82 (2024)
doi:10.1007/s41365-024-01446-w

\bibitem{Li:2023}
Y.~Li, C.~Li, K.~Hu, et al.,
Design and development of multi-channel front end electronics based on dual-polarity charge-to-digital converter for SiPM detector applications,
Nucl. Sci. Tech. \textbf{34}, no.2, 18 (2023)
doi:10.1007/s41365-023-01168-5

\bibitem{zhao2024}
Z.~Y.~Zhao, B.~H.~Qi, S.~Li, et al.,
Dynamic Range of SiPMs with High Pixel Densities. 
[arXiv:2407.17794 [physics.ins-det]]

\bibitem{S14160-3010PS}
HAHAMATSU, 
S4160-3010PS. 
\url{https://www.hamamatsu.com.cn/cn/zh-cn/product/optical-sensors/mppc/mppc_mppc-array/S14160-3010PS.html}

\bibitem{kotera2016describing}
K.~Kotera, W.~Choi, T.~Takeshita,
Describing the response of saturated SiPMs. 
[arXiv:1510.01102 [physics.ins-det]]

\bibitem{PIEMONTE20192}
C.~Piemonte and A.~Gola,
Overview on the main parameters and technology of modern Silicon Photomultipliers. 
Nucl. Instrum. Meth. A \textbf{926}, 2--15 (2019). 
doi:10.1016/j.nima.2018.11.119

\bibitem{NAGY201444}
F.~Nagy, M.~Mazzillo, L.~Renna, et al.,
Afterpulse and delayed crosstalk analysis on a STMicroelectronics silicon photomultiplier. 
Nucl. Instrum. Meth. A \textbf{759}, 44--49 (2014). 
doi:10.1016/j.nima.2014.04.045

\bibitem{Zhu:2019ihr}
R.~Y.~Zhu,
Ultrafast and Radiation Hard Inorganic Scintillators for Future HEP Experiments. 
J. Phys. Conf. Ser. \textbf{1162}, 012022 (2019)
doi:10.1088/1742-6596/1162/1/012022

\bibitem{ULYANOV2020164203}
A.~Ulyanov, D.~Murphy, J.~Mangan, et al.,
Radiation Damage Study of SensL J-Series Silicon Photomultipliers Using 101.4 MeV Protons. 
Nucl. Instrum. Meth. A \textbf{976}, 164203 (2020)
doi:10.1016/j.nima.2020.164203

\bibitem{SANCHEZMAJOS2009506}
S.~Sanchez Majos, P.~Achenbach, C.~Ayerbe Gayoso, et al.,
Noise and radiation damage in silicon photomultipliers exposed to electromagnetic and hadronic radiation. 
Nucl. Instrum. Meth. A \textbf{602}, 506--510 (2009)
doi:10.1016/j.nima.2009.01.176

\bibitem{PREGHENELLA2023167661}
R.~Preghenella, M.~Alexeev, P.~Antonioli, et al.,
A SiPM-based optical readout system for the EIC dual-radiator RICH. 
Nucl. Instrum. Meth. A \textbf{1046}, 167661 (2023)
doi:10.1016/j.nima.2022.167661

\end{thebibliography}
\end{document}